\documentclass{article}
\usepackage[english]{babel}
\usepackage[letterpaper, top=2cm, bottom=2cm, left=3cm, right=3cm, marginparwidth=1.75cm]{geometry}
\usepackage[T1]{fontenc}
\usepackage{lmodern}
\usepackage{upgreek}
\usepackage[labelfont=bf]{caption}
\usepackage{floatrow}
\usepackage{setspace}
\usepackage{amsmath}
\usepackage[nice]{nicefrac}
\usepackage{graphicx}
\usepackage{float}
\usepackage[colorlinks=false, hidelinks]{hyperref}
\usepackage[shortlabels]{enumitem} 
\usepackage{changepage}
\usepackage{gensymb}
\usepackage[pdftex]{color}
\usepackage{titling}
\usepackage[affil-it]{authblk}
\usepackage{hyperref}
\hypersetup{
    pdftitle={Rotary Excitation of peak-like magnetic field fluctuations},
    pdfauthor={Petra Albertova},
    pdfsubject={Preprint},
    pdfproducer={}
}

\title{\textbf{Rotary Excitation of non-sinusoidal pulsed magnetic fields: Towards non-invasive direct detection of cardiac conduction}}
\vfill
\author[1,2,$\star$]{Petra Albertova}
\author[1,2,$\star$]{Maximilian Gram}
\author[3]{Martin Blaimer}
\author[1]{Wolfgang R. Bauer}
\author[2]{Peter M. Jakob}
\author[1]{Peter Nordbeck}

\affil[1]{\normalsize Department of Internal Medicine I, University Hospital Würzburg, Würzburg, Germany\vspace*{0.1cm}}
\affil[2]{\normalsize Experimental Physics 5, University of Würzburg, Würzburg, Germany\vspace*{0.1cm}}
\affil[3]{\normalsize Fraunhofer Institute for Integrated Circuits IIS, Würzburg, Germany\vspace*{0.1cm}}
\affil[$\star$]{\normalsize these authors contributed equally to this work \vspace*{0.1cm}}

\date{}
\begin{document}
\begin{titlepage}
\clearpage 
\setlength{\droptitle}{-1.5cm} 
\maketitle
\thispagestyle{empty}
\vspace{-1.2cm}
\noindent
\noindent
\begin{spacing}{1.115}
\begin{tabular}{l l}
\textbf{Address for correspondence:} & Petra Albertová, Experimental Physics 5,\\
 & Faculty of Physics and Astronomy,\\
 & University of Würzburg\\
 & petra.albertova@uni-wuerzburg.de\\
 & 
\end{tabular}\\
\noindent
\begin{tabular}{l l l}
\textbf{Author’s ORCID:} & Petra Albertova: & \href{https://orcid.org/0000-0003-3646-7019}{orcid.org/0000-0003-3646-7019}\\
& Maximilian Gram: & \href{https://orcid.org/0000-0003-2184-3325}{orcid.org/0000-0003-2184-3325}\\
& Martin Blaimer: & \href{https://orcid.org/0000-0002-6360-9871}{orcid.org/0000-0002-6360-9871}\\
& Wolfgang R. Bauer: & \href{https://orcid.org/0000-0002-4652-7313}{orcid.org/0000-0002-4652-7313}\\
& Peter M Jakob: & \href{https://orcid.org/0000-0002-2904-2576}{orcid.org/0000-0002-2904-2576}\\
& Peter Nordbeck: & \href{https://orcid.org/0000-0002-2560-4068}{orcid.org/0000-0002-2560-4068}
\end{tabular}\\

\noindent
\section*{Abstract}
\vspace{-0.2cm}
\noindent
\textbf{Purpose:} 
In the recent past, spin-locking MRI experiments were successfully applied for the direct detection of sinusoidal magnetic field oscillations in the sub-nT range. In the present study, this detection approach was extended to non-sinusoidal pulsed magnetic fields based on the Rotary Excitation (REX) mechanism.\\
\textbf{Methods:} 
The new detection concept was examined by means of Bloch simulations, evaluating the interaction effect of spin-locked magnetization and low-frequency pulsed magnetic fields. The REX detection approach was validated under controlled conditions in phantom experiments at 3\,T. Gaussian and Sinc-shaped stimuli were investigated. In addition, the detection of artificial fields resembling a cardiac QRS complex, which is the most prominent peak visible on a Magnetocardiogram, was tested.\\
\textbf{Results:} 
Bloch simulations demonstrated that the REX method has a high sensitivity to pulsed fields in the resonance case, which is met when the spin-lock frequency coincides with a non-zero Fourier component of the stimulus field. In the experiments, we found that magnetic stimuli of different durations and waveforms can be distinguished by their characteristic REX response spectrum. The detected REX amplitude was proportional to the stimulus peak amplitude ($R^2\,>\,0.98$) and the lowest field detection was 1\,nT. Furthermore, the detection of QRS-like fields with varying QRS durations yielded significant results in a phantom setup ($p\,<\,0.001$).\\
\textbf{Conclusion:}\,
REX detection can be transferred to non-sinusoidal pulsed magnetic fields and could provide a non-invasive, quantitative tool for spatially resolved assessment of cardiac biomagnetism. Potential applications include the direct detection and characterization of cardiac conduction.\\
\noindent
\textbf{Key words:} 	spin-lock, rotary excitation, REX, biomagnetism, $T_{\text{1}\uprho}$, MCG\\
\begin{center}
\Large{\textbf{Submitted to Magnetic Resonance in Medicine}}
\end{center}
\end{spacing}
\end{titlepage}
\begin{spacing}{1.115}

\newpage
\section*{Introduction}
For basic physiological and medical research, varying biomagnetic fields intrinsically occurring in the human body hold great interest, since electromagnetic signals are of central importance for transmission of information in living organisms [1]. Therefore, measuring bioelectric or closely related biomagnetic fields allows conclusions to be drawn about the functionality of the involved systems. Examples include neuronal and cardiac fields, which are so far measured by Magnetoencephalography (MEG) or Magnetocardiography (MCG) [2, 3, 4, 5]. Although these modalities realize detections in the pT range, a major drawback is the limited spatial resolution compared to MRI. For spatially resolved detection via MRI, the sensitivity as well as the low-frequency nature of biomagnetic activity has been the major challenge so far [6]. Since the sensitivity of clinical MRI scanners is fixed by the strength of the main magnetic field to processes in the MHz range, a mechanism of interaction other than resonant excitation in a quantum mechanical two-level system needs to be employed. Moreover, techniques that detect static fields via gradient echoes are unsuitable because time-varying fields might lead to cancellation of the observed phase evolution [7].\\
In 2008, an innovative approach for MRI based detection of oscillating fields was published by Witzel et al. [8]. The main idea is to shift sensitivity to another frequency by applying a spin-lock (SL) preparation. A SL pulse is a resonant continuous wave pulse, which is applied parallel to the transverse magnetization [9, 10]. In this condition, spin-locking introduces a second resonance frequency adjustable by the SL field strength. If the frequency of a field oscillation (stimulus) matches the SL frequency, resonant energy absorption induces a torsion of the spin-locked magnetization [11, 12]. This phenomenon is called Rotary EXcitation (REX) and was trialed for imaging alpha activity [13]. Recently, REX has been validated to yield in vivo sensitivity down to at least 1\,nT [14].\\
To this point, all proposed methods were designed in view of sinusoidal oscillatory stimuli [8, 13, 15], since neural fields in activated cortical regions arise from periodic activation at cellular level [16]. Thus, the original detection concept presented by Truong et al. was based on the following principles [13]:
\begin{enumerate}[I]
\item{Oscillatory biomagnetic stimuli act like periodic gyromagnetic forces in the SL state.}
\item{Rotary excitation arises if the frequency of a stimulus matches the SL frequency.}
\item{The resulting REX prepared magnetization depends on the relative phase between SL and stimulus.}
\end{enumerate}
Although remarkable sensitivities have been demonstrated in previous studies [11, 15, 17], so far only pure sinusoidal fields were detected, which severely limits possible applications. A generalization of the REX method towards detection of magnetic waveforms of arbitrary shape, including pulsated, peak-like fields would open up a wide range of applications for other sources of biomagnetism. For example, individual axons [18] and nerves [19], which exhibit a bipolar course of the magnetic amplitude, or peak-like fields of activated brain regions [20] might be examined. However, the probably most valuable application for basic research and clinical diagnostics is non-invasive sensing of cardiac biomagnetism [21]. In this case, there is enormous need, as the assessment of cardiac conduction by means of ECG and MCG does not offer sufficient spatial resolution [22] and invasive catheter-based techniques are slow and a burden for patients [23, 24]. In addition, cardiac conduction has been extensively researched, as the aforementioned methods provide rough estimates of the fields to be expected [25], which makes the development of a dedicated REX method conceivable.\\
In the current work, we have extended the concept of REX detection to achieve direct sensing of pulsed magnetic fields (PMF). As the main goal, we focused on cardiac biomagnetism, whereby QRS-shaped fields need to be detected. For this purpose, we generalized the three postulates mentioned above and optimized our new concept using Bloch simulations. We subsequently validated the detection of basic Gaussian and Sinc-shaped fields on a clinical MRI scanner and conducted the first detections of QRS-like fields in tissue-like phantoms.

\newpage
\section*{Methods}
In the following section, we first explain the concept of REX based PMF detection. We focus on the detection of non-sinusoidal magnetic fields, such as Gaussian and Sinc-stimuli as well as QRS-like fields, which represent the most prominent peaks visible on both ECG and MCG. Subsequently, the Bloch simulations and the phantom experiments are described. All measurements were performed on a clinical 3\,T scanner (MAGNETOM Skyra, Siemens Healthineers, Erlangen, Germany).

\subsection*{Concept of REX detection}
The concept of SL based field detection of pure sinusoidal fields used as a starting point is explained in [14]. A magnetic stimulus with frequency $f_{\text{stim}}$ induces rotary excitation of transverse magnetization if a SL pulse with a field amplitude $B_{\text{SL}}$ matching the stimulus frequency according to $f_{\text{SL}}\,=\,f_{\text{stim}}\,=\,\gamma / 2\pi\,B_{\text{SL}}$ is present. Here, $\gamma$ denotes the gyromagnetic ratio. In this resonant case, the magnetization, previously initialized as spin-lock component on the $x'$ axis, is partially converted to spin-tip components in the $y'z'$ plane. A detection is performed by imaging the $z'$ projection. For the transfer of REX detection to non-sinusoidal, pulsed stimuli, we set up three postulates:
\begin{enumerate}[I]
\item Pulsed biomagnetic stimuli act like punctual gyromagnetic forces in the SL state.
\item Rotary excitation arises if the stimulus has a significant component in Fourier space matching the SL frequency.
\item The resulting REX prepared magnetization depends on the relative timing of the interaction between SL and stimulus.
\end{enumerate}
Similar to the original REX experiment employing a sinusoidal stimulus, field detection is attributed to an observation of varying $z'$ components [11, 12, 13]. Thus, multiple measurements are required, with the relative timing between stimulus and SL being varied (Fig. 1). In this case, we expect a sinusoidal course of the REX signal $S_{\text{REX}}$ with regard to the interaction timing and a dependence of the effect on the adjusted SL amplitude. Accordingly, either the amplitude of the observed signal or its standard deviation can be evaluated for detection.
\subsection*{Bloch simulation of REX detection}
The proposed detection concept was investigated by means of Bloch simulations. For the simulation of the interaction in the SL state, the following model was considered, assuming a SL pulse in the $x'$ direction of the rotating reference frame [26]:
\begin{equation}
\frac{\text{d}}{\text{d}t} \vec{M}'\,=\,\left( \begin{array}{ccc}
-1/T_{1\uprho} & \gamma \Delta B_0^{\text{stim}}(t) & 0 \\
-\gamma \Delta B_0^{\text{stim}}(t) & -1/T_{2\uprho} & 2\pi f_{\text{SL}} \\
0 & -2\pi f_{\text{SL}} & -1/T_{2\uprho} \\
\end{array} \right)\,\cdot\,\vec{M}'
\end{equation}
Here, $T_{1\uprho}$ and $T_{2\uprho}$ are the rotating frame relaxation times and $\Delta B_0^{\text{stim}}(t)$ is the time-dependent stimulus amplitude. The solution to the system of nonlinear differential equations for the magnetization trajectory $\vec{M}'(t)$ was determined numerically by Runge-Kutta integration (4th order, time steps $1\,\cdot\,10^{-5}$\,s) in Matlab (R2022a, The MathWorks Inc., Natick, Massachusetts, USA). For simulating REX detection, the $M_{z'}$ magnetization was calculated over different interaction timings. The standard deviation of the simulated signal was used as the measure of detection and is referred to as the REX amplitude $A_{\text{REX}}$ as introduced in [14].\\
In order to evaluate the postulates (I-III) for REX detection of PMFs, two peak-like Gauss and Sinc-shaped stimuli were examined besides a sinusoidal field (Fig. 2A). Gaussian and Sinc-stimuli rely on simple mathematical modeling and were calculated analogously to the envelope of RF waveforms [27, 28]. Here, a duration of $\tau_{\text{stim}}\,=\,50$\,ms was used. To diversify shape and spectral characteristics, PMFs with different time-bandwidth products ($t\,\cdot\,BW\,=\,4\,\ldots\,10$) were considered. REX detection was simulated for different SL amplitudes to investigate the spectral response. For this purpose, $f_{\text{SL}}$ was varied in the range $4\,\ldots\,200$\,Hz (step size of $0.5$\,Hz). In addition, the spectral power of the fields was calculated via Fourier transform to validate postulate II. To consider the impact of relaxation, simulations were performed with and without consideration of relaxation (Eq. 1).
\begin{figure}[H]
\caption*{\textbf{Figure 1)} REX detection concept for pulsed magnetic fields. The stimulus interacts with the magnetization after it was aligned along the $x'$ axis by a 90° tip-down pulse. During interaction, a SL pulse shifts sensitivity into the low frequency range for its pulse duration $t_{\text{SL}}$. The SL duration is the result of two buffer times, each equal to half the pulse duration $\tau_{\text{stim}}$, and an interaction time period $t_{\text{inter}}$. This period indicates the interval in which the stimulus peak must be varied to achieve a phase variation of $2\pi$ according to Eq. 4. The buffer time ensures that each stimulus remains entirely within $t_{\text{SL}}$. \textbf{A)} If the SL pulse and stimulus are resonant, rotary excitation causes a significant deflection of the magnetization and thereby the emergence of a non-vanishing $M_z$ component. The final component depends on the timing of interaction and its variation over various interactions is used for detection (visualization in Supporting Video 1). \textbf{B)} An offresonant interaction does not lead to significant formation of an $M_z$ component.}
\includegraphics[width=0.75\textwidth]{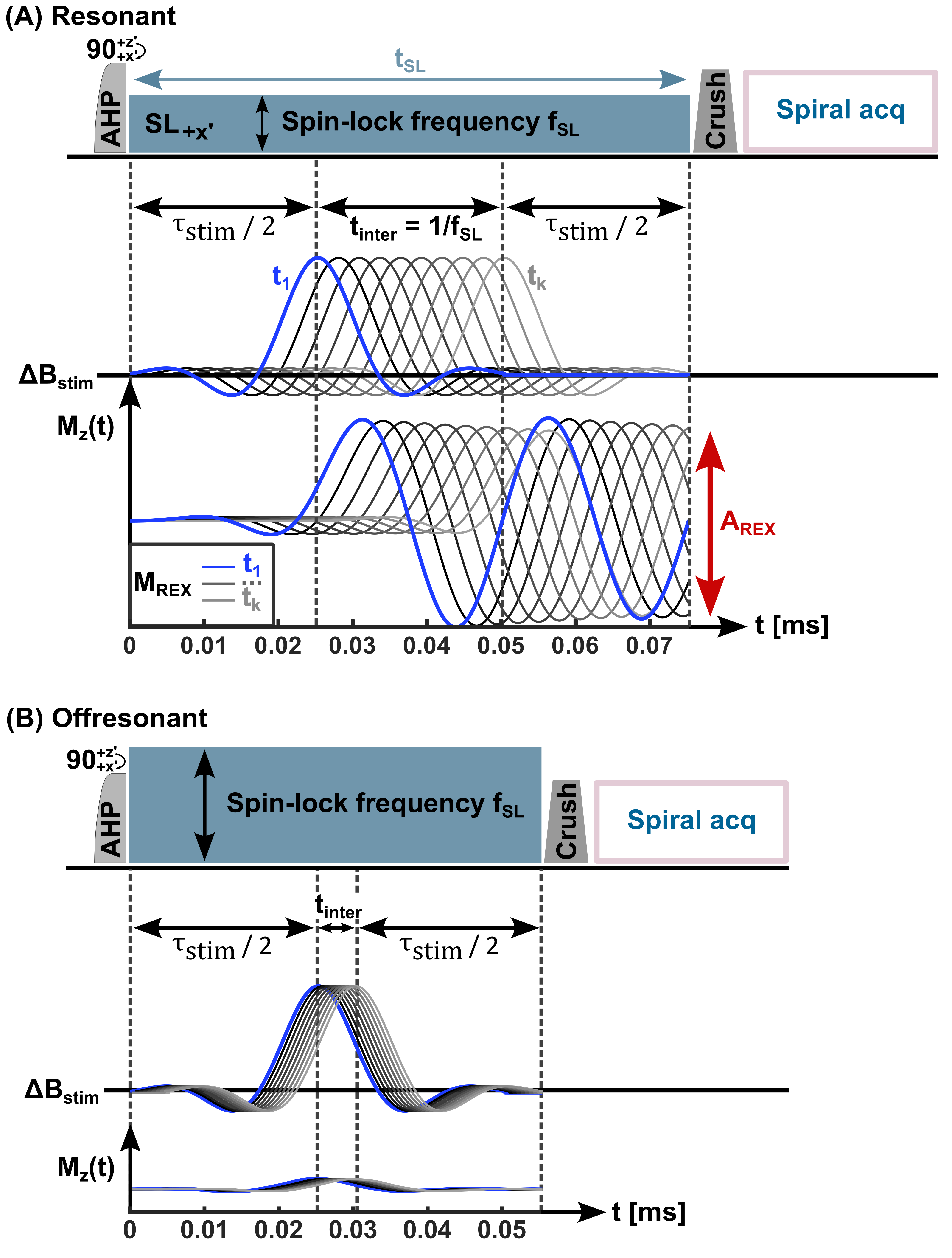}
\end{figure}

In order to obtain reasonable interaction timings for detection of PMFs with durations $\tau_{\text{stim}}\,=\,50$\,ms, we employed the following preparation times $t_{\text{SL}}$ (Fig. 1):
\begin{equation}
t_{\text{SL}}\,=\,\tau_{\text{stim}}\,+\,\frac{1}{f_{\text{SL}}} 
\end{equation}
The SL pulse is therefore longer than the stimulus to be detected. The interval $1/f_{\text{SL}}$ is used for varying the timing $t_{\text{inter}}$ of field interaction relative to the SL pulse and provides possible interaction timings causing a full $2\pi$ phase variation. In analogy to a sinusoidal stimulus, the interaction time 
\begin{equation}
t_{\text{inter}}\,=\,\frac{1}{2}\,\cdot\,\tau_{\text{stim}}\,+\,\frac{\phi}{2\pi}\,\cdot\,\frac{1}{f_{\text{SL}}} 
\end{equation}
can be expressed by a phase shift:
\begin{equation}
\phi\,=\,f_{\text{SL}}\,\cdot\,\left( t_{\text{inter}}\,-\,\frac{\tau_{\text{stim}}}{2} \right)
\end{equation}
\begin{figure}[htbp]
\caption*{\textbf{Figure 2)} Pulsed field fluctuations cause rotary excitation of spin-locked magnetization in the resonance case. \textbf{A)} Exemplary stimulus waveforms. The Figure depicts a Sine-function, a Gaussian peak and a truncated Sinc. \textbf{B)} Magnetic oscillations have a sharply focused distribution in their power spectral density spectrum. In contrast, pulsed magnetic fields are composed of broad frequency spectra. \textbf{C)} The Bloch equations predict rotary excitation for all pulsed fields investigated. The simulation of response spectra shows a curve qualitatively corresponding to the power spectral density confirming the anticipated resonance effect. Since no noise is considered in the simulation, a vanishing REX amplitude results if the REX effect does not occur. The REX amplitude was determined by calculating the standard deviation over ten interaction timings for each SL frequency.}
\includegraphics[width=0.75\textwidth]{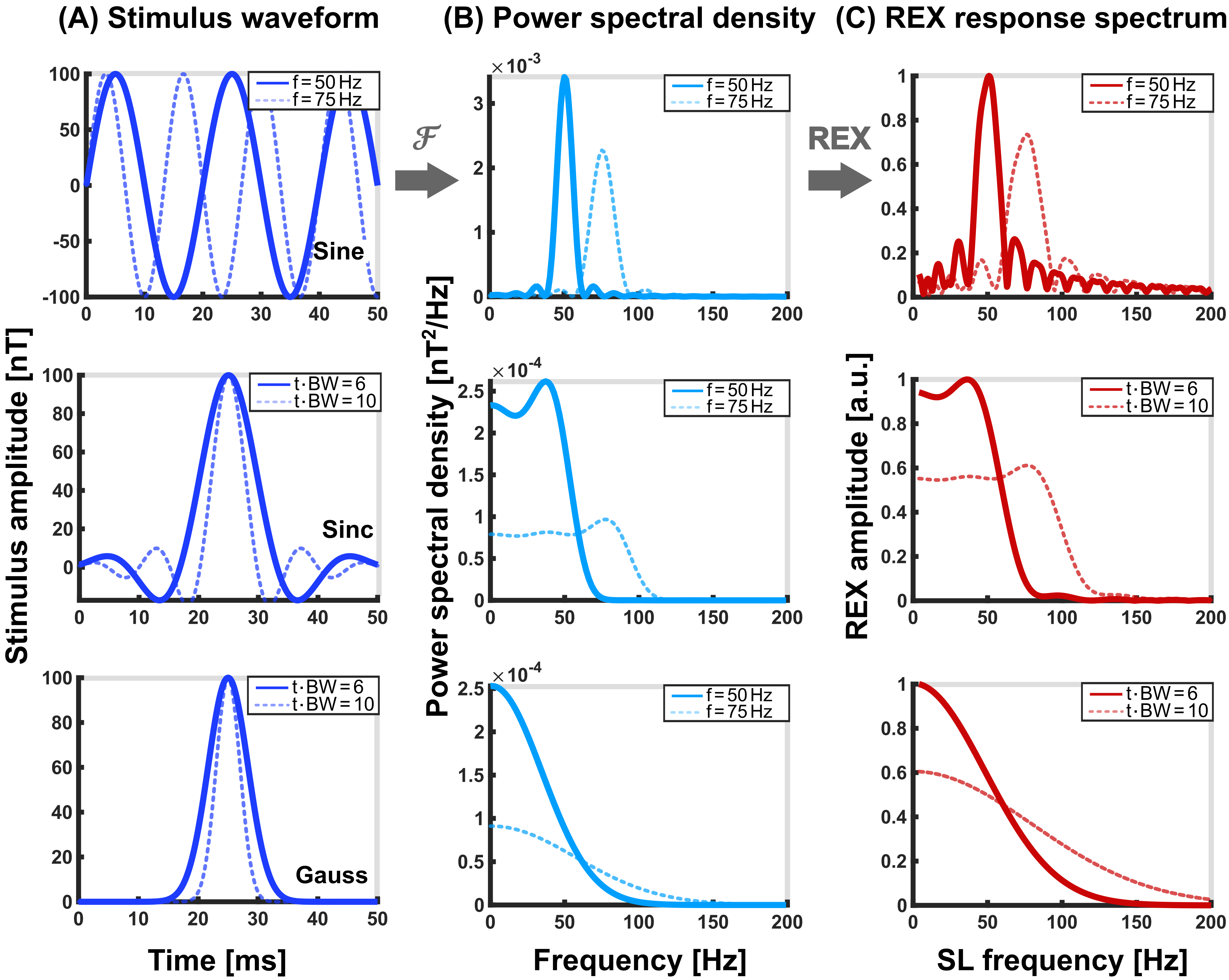}
\end{figure}

\subsection*{Experimental validation: Detection of Gauss and Sinc-shaped fields}
For experimental validation, a spherical calibration phantom (diameter $17.5$\,cm) of demineralized water doped with $1.25$\,g/L NiSO$_4$ was used. A SL preparation module followed by interleaved spiral readouts for image acquisition ($4$ interleaves, $TR\,=\,1$\,s, $N_x\,=\,N_y\,=\,128$, $FOV\,=\,240\,\times\,240$\,mm\textsuperscript{2}) was implemented in Pulseq [27]. The sequence diagram is illustrated in Fig. 1. The magnetization is initialized by an adiabatic-half-passage (AHP) $90^\circ$ excitation pulse ($3$\,ms). Adiabatic pulses are used for this purpose in order to ensure uniform excitation unaffected by $B_1^+$ inhomogeneities [29]. The PMF waveforms $\Delta B_0^{\text{stim}}(t)$ were generated as proposed in [14] using the concept of transmission of Rotary EXcitation (tREX) by the built-in gradient system (max. gradient $45$\,mT/m, max. slew rate $200$\,T/m/s) in an offcenter slice ($|\Delta z|\,=\,15$\,mm). Crusher gradients were applied after tREX interaction to dephase remaining transverse magnetization and to achieve imaging of the $z'$ magnetization. Signal acquisition was carried out with a $20$-channel receive head-neck array.\\
In order to verify the occurrence of the REX effect in case of an interaction with a PMF (postulate I), first experiments with a Sinc-shaped stimulus with duration $\tau_{\text{stim}}\,=\,50$\,ms, $t\,\cdot\,BW\,=\,6$ and peak amplitude $100$\,nT were performed. The SL pulse duration was $t_{\text{SL}}\,=\,75$\,ms. The SL pulse amplitude was set to $f_{\text{SL}}\,=\,40$\,Hz. This choice was based on the analysis of the power spectral density of the stimulus, which shows maximum values for $40$\,Hz (Fig. 2B). Consequently, we expect to observe the largest REX effect for this parameter choice fulfilling resonance condition (postulate II). For detection, $10$ different interaction timings $t_{\text{inter}}$ in the range $25\,\ldots\,50$\,ms were investigated and $A_{\text{REX}}$ was determined from the standard deviation of the measured signals (Fig. 1). Furthermore, in analogy to the simulations, the REX response spectrum was measured by performing a variation of the SL amplitude for each stimulus type. For this purpose, REX scans with $f_{\text{SL}}$ in the range $8\,\ldots\,200$\,Hz (step size $8$\,Hz) were conducted. The full scan time for each spectrum was $\approx\,19$\,min. In order to obtain a result that is minimally affected by static $\Delta B_0$ and $\Delta B_1^+$ deviations, field maps were acquired using the WASABI method [30]. Based on these field maps, a region with low field deviations ($|\Delta f_0|\,<\,5$\,Hz, $|\Delta B_1^+|\,<\,5\,\%$) was selected for evaluation of the REX signal. The measured REX spectra were compared with the power spectral density of the respective Fourier transforms, as well as with the Bloch simulation results. 

\subsection*{Experimental application: Detection of QRS-like fields in a phantom}
In a second phantom experiment, REX detection was applied for the assessment of QRS-like fields. The QRS complex corresponds to the left and right ventricular depolarization of the heart and is the strongest source of biomagnetism visible in MCGs [25]. It indicates the propagation of an electrical impulse through the conducting Purkinje fibers and ventricular contraction [31], consisting of a negative deflection (Q) and a central peak (R) followed by a dip (S). The QRS duration in humans ranges from $80$ to $120$\,ms [32] and the magnitude of the R-wave is estimated to be about $14$\,nT locally inside of the heart [33].\\
In order to demonstrate REX detection of such fields, experiments were performed in a myocardial-tissue-like BSA (Bovine Serum Albumin, Sigma-Aldrich, St. Louis, MO, USA) phantom with relaxation times $T_{1\uprho}\,=\,65$\,ms and $T_{2\uprho}\,=\,90$\,ms comparable to literature estimates [34, 35]. Artificial fields were generated via tREX analogous to the measurements of Gauss and Sinc peaks. The phantom with thermally cross-linked BSA (concentration $15\,\%$) was inserted into a water-filled container with a NaCl concentration of $5$\,g/L. The QRS-like stimuli were modeled from MCG data sets [36]. The aim of the experiments was to emulate in vivo conditions as closely as possible. However, a major challenge for an in vivo sequence arises from the fact that the relative timing between SL and R-wave cannot be adjusted but is dictated by the natural variation of the cardiac cycle. To adequately model this condition, an ECG dataset from a healthy volunteer was analyzed, which was recorded during a conventional cardio MRI setting (breath hold of $15$\,s approximately every $2$\,min). A total of $526$ heartbeats were analyzed to model natural RR variation.\\
Fig. 3A illustrates the concept of the developed QRS detection sequence. The sequence is based on a prospective trigger leading to an SL interaction of every second QRS complex. The fixed delay after triggering was adjusted so that the R-wave exactly hits the middle of the SL pulse for an RR interval of $850$\,ms. For an RR below ($\Delta \text{RR}\,<\,0$) or above ($\Delta \text{RR}\,>\,0$) $850$\,ms, the R-wave is shifted left or right relative to the SL pulse. A detection was performed with $20$ interactions, again evaluating the standard deviation of the signal. RR intervals for which the shift was so large that they corresponded to a relative phase of more than $\pm\,\pi$ were not measured. After exclusion a set of $20$ interaction timings was used for detection. On the ECG data set used (Fig. 3B), this criterion led to an exclusion of $45.4\,\%$. Imaging was performed in the light of an in vivo experiment using a fast spiral readout, which achieves full k-space sampling after each SL preparation with four interleaved spirals and ramped flip angles ($\alpha\,= \,25^\circ, 27.8^\circ, 31.8^\circ, 38.3^\circ$; $N_x\,=\,N_y\,=\,128$; $FOV\,=\,240\,\times\,240$\,mm\textsuperscript{2}).\\
\begin{figure}[htbp]
\caption*{\textbf{Figure 3)} Concept for the detection of cardiac magnetic fields. \textbf{A)} The R-wave of an ECG signal is used to trigger the sequence. The trigger starts a fixed delay in the sequence that is adapted to the patient so that on average the R-wave of the next QRS complex coincides with the center of the SL pulse ($\Delta \text{RR}\,=\,0$). However, due to the natural variation of the RR interval, the exact interaction time is randomly distributed. \textbf{B)} Variation of the RR interval of a healthy volunteer obtained from ECG data under breath instructions. Data with a variation $\Delta \text{RR}$ corresponding to $\phi\,>\,\pm\,\pi$ are retrospectively discarded ($|\Delta \text{RR}|\,<\,0.5\,\cdot\,1/f_{\text{SL}}\,=\,0.5/12\,$Hz). \textbf{C)} Exemplary interactions of QRS fields with the SL pulse. The QRS waveforms are cropped, since the optimized SL pulse duration ($t_{\text{SL}}\,=\,73$\,ms) is shorter than the full QRS duration.}
\includegraphics[width=0.75\textwidth]{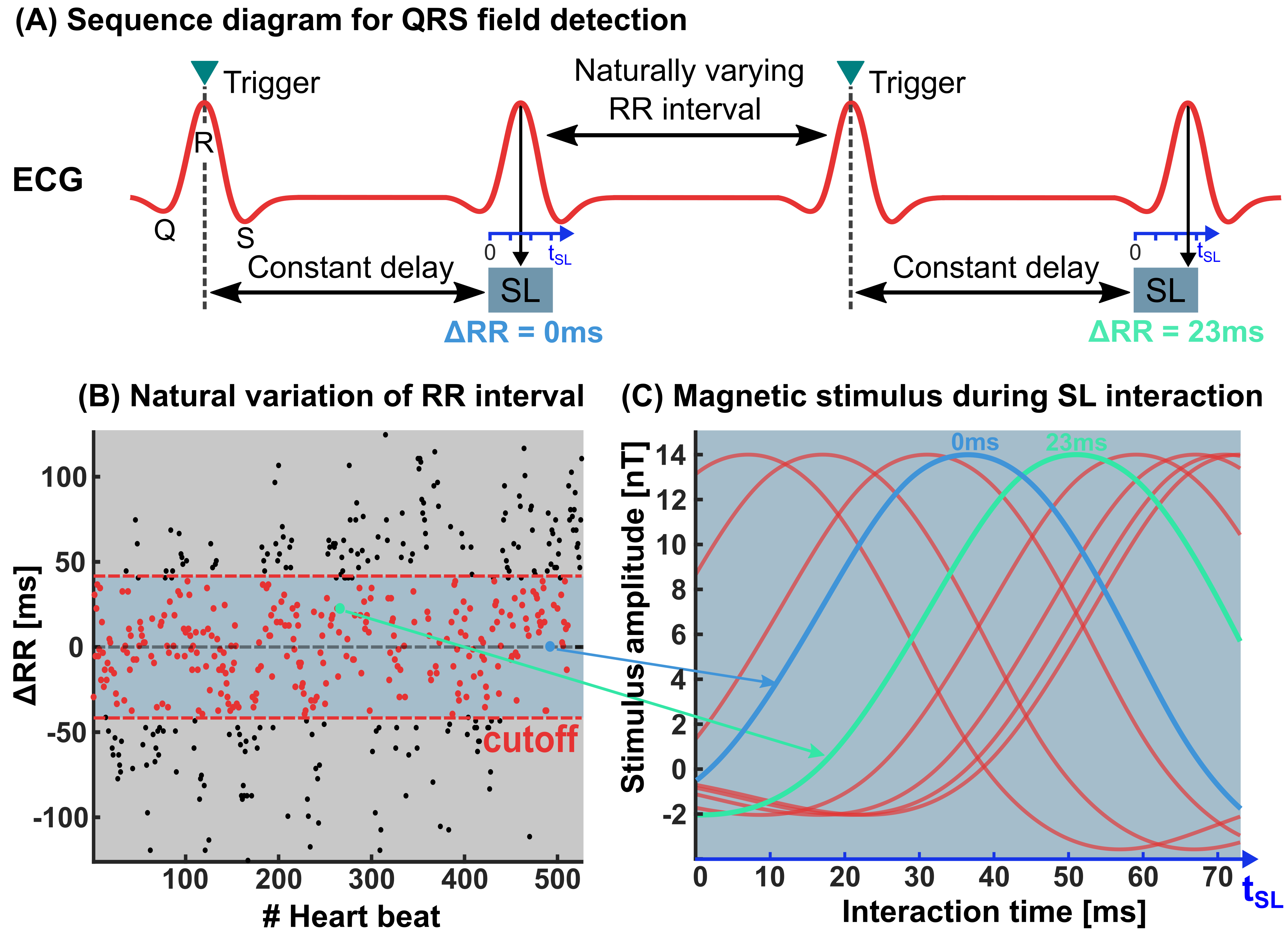}
\end{figure}\\
\noindent
Prior to the first experiment, the parameters of the SL preparation ($t_{\text{SL}}$ and $f_{\text{SL}}$) were optimized by means of a Bloch simulation. A QRS duration of $100$\,ms and a magnitude of $14$\,nT were assumed. For the $T_{1\uprho}$ and $T_{2\uprho}$ relaxation, the values $50$\,ms and $70$\,ms were adopted respectively, which correspond to reasonable values for myocardial tissue [34, 35].\\
The first part of the experimental validation consisted of attempting REX detection for a stimulus with $100$\,ms QRS duration for peak amplitudes in the range $1\,\ldots\,15$\,nT. It was examined whether the measured REX amplitude correlates with the peak amplitude. For this purpose, the optimized SL parameters according to the simulation results ($f_{\text{SL}}\,=\,12$\,Hz, $t_{\text{SL}}\,=\,73$\,ms) were applied, considering the previously experimentally determined $B_1^+$ deviation of $123\,\%$ averaged within the BSA phantom using WASABI [30] ($f_{\text{SL}}^{\text{set}}\,=\,12/1.23\,\text{Hz}\,=\,9.76$\,Hz). Since in vivo not only the RR intervals but also the QRS durations underlie physiological variations and vary among individuals, it was investigated in the second part of the experiment whether QRS fields with $80$, $90$, $100$, $110$ and $120$\,ms can be significantly detected at the expected peak magnitude of $14$\,nT. For this purpose, the detection was repeated ten times for each QRS duration. The respective samples of the determined $A_{\text{REX}}$ values were compared voxel-wise with a control sample measured at an SL frequency of $100$\,Hz (i.e. instead of $f_{\text{SL}}\,=\,12$\,Hz $f_{\text{SL}}^{\text{set}}\,=\,100/1.23\,\text{Hz}\,=\,81.3$\,Hz was used). In this control group, the REX effect is expected to be eliminated since SL and stimulus frequency do not match. A two-sample t-test was performed for each voxel within the BSA tube and the corresponding $p$-values were calculated as a measure for significant detection. Another detection experiment was performed at $f_{\text{SL}}\,=\,12$\,Hz and $0$\,nT. However, since the QRS complex cannot be "switched off" during an in vivo experiment, the offresonant sample was used as the control and the $0$\,nT sample was only considered as an additional validation. The measurement time of a single sample was $6.2$\,min, which resulted in the total acquisition of all seven detection series taking approximately $44$\,min. In order to reduce drift effects during the lengthy experiment, adjustments (shim and basic frequency) were repeated prior to each sample.

\newpage
\section*{Results}
\subsection*{Bloch simulation of REX detection}
Figure 1 illustrates the simulated magnetization component $M_{z'}$ for the interaction with a Sinc-stimulus. It can be recognized that the PMF acts like a punctual force and causes a nutation around the SL axis (Fig. 1A, postulate I). It was further visualized what effect the identical stimulus produces at a significantly higher $f_{\text{SL}}$. Here, approximately zero $M_{z'}$ components are obtained, indicating that REX resonance condition was not met (Fig. 1B, postulate II). For REX resonance, the timing of the interaction during spin-locking determines the final prepared magnetization and the prepared $M_{z'}$ component exhibits a sinusoidal response (postulate III). This phenomenon was animated in the Supporting Material (Supporting Video 1).\\
In Figure 2, the signal waveforms and their corresponding power spectral densities are depicted for the different stimulus types and compared with simulated REX responses. This verifies postulate II, since there is a high correlation between REX response and the power spectral density. The FWHM values, indicating the frequency range in which the resonance condition is met, are $\approx\,40$\,Hz (Gauss) and $\approx\,50$\,Hz (Sinc) for $\tau_{\text{stim}}\,=\,50$\,ms at $t\,\cdot\,BW\,=\,6$. The global maximum of Sinc is $\approx\,40$\,Hz. If relaxation effects are additionally considered in the Bloch simulation (Fig. 4), the linear correlation does not hold for all $f_{\text{SL}}$. Particularly for long $t_{\text{SL}}$, which were used for small $f_{\text{SL}}$ (Eq. 2), the impact of relaxation is clearly noticeable, and REX spectra deviate significantly from Fourier spectra.
\begin{figure}[htbp]
\caption*{\textbf{Figure 4)} REX response spectra for Sinc and Gauss stimuli. \textbf{A)} Sinc and \textbf{B)} Gauss stimuli were simulated without considering relaxation (black, $T_{1\uprho}$ and $T_{2\uprho}$ were assumed to be infinity) and with relaxation times corresponding to the values of the spherical phantom (gray) and myocardial tissue (light gray). A decrease in the REX amplitude due to relaxation is particularly pronounced at low frequencies, since according to Eq. 2 the SL duration was extended in this range. However, the response pattern for differing stimuli remains both distinctive and distinguishable.}
\includegraphics[width=0.85\textwidth]{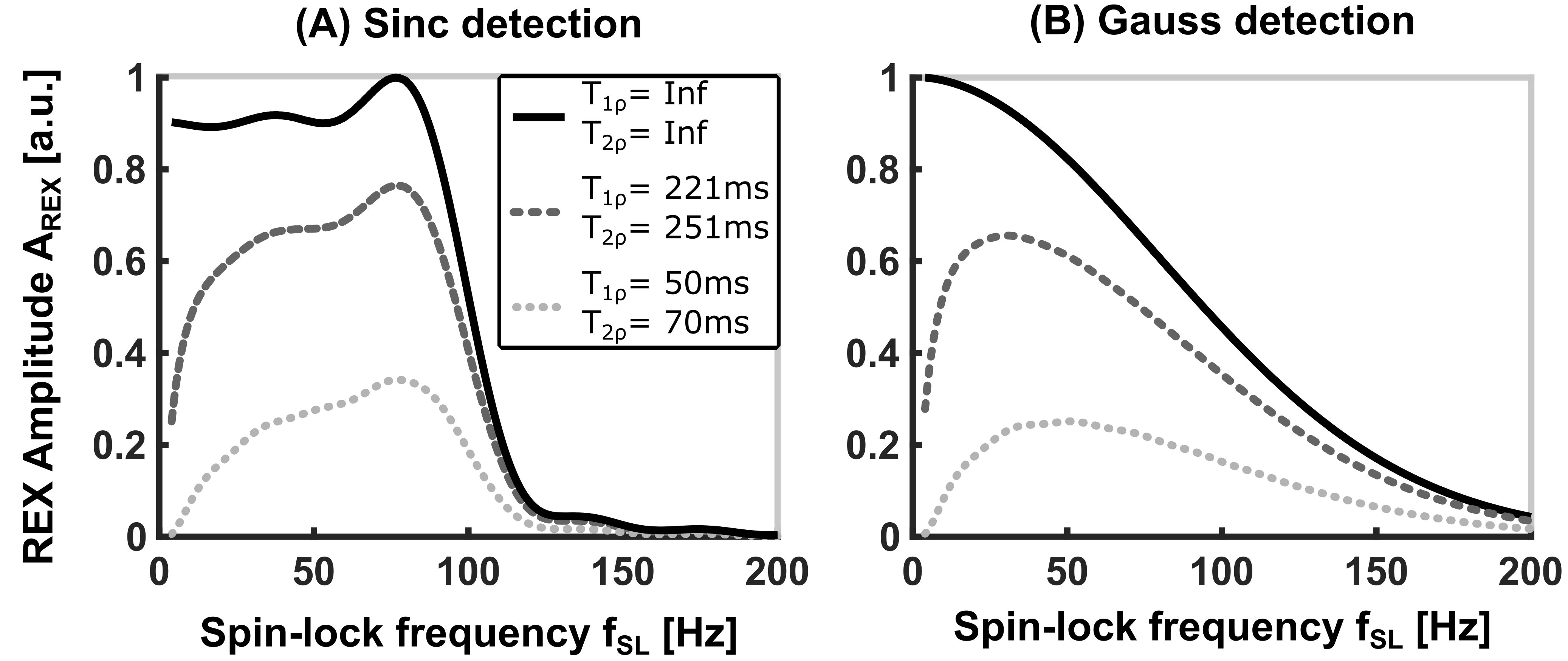}
\end{figure}

\subsection*{Detection of Gauss and Sinc-shaped fields}
The acquired REX based contrast images of Sinc-detection are depicted in Figure 5. A sinusoidal variation of the REX contrast can be observed. If the interaction time varies by $1/f_{\text{SL}}$, which corresponds to a relative phase shift $\Delta \phi\,=\,2\pi$ (Eq. 3), the REX signal traverses a full period (Fig. 5B). In the phantom experiment, the time of interaction is precisely known and adjustable. Hence, it is possible to use a sinusoidal function to perform regression analysis of the contrast images.
\begin{equation}
S_{\text{REX}}\,=\,A_{\text{REX}}^{\text{fit}}\,\cdot\,\sin(\phi\,+\,\phi_0) 
\end{equation}
Before calculating the voxel-wise fit, an offset correction was performed by subtracting the complex mean of the REX signal in each voxel. The resulting $A_{\text{REX}}$ map has a high structural similarity with the standard deviation map scaled by the factor $\sqrt{2}$. Yet, the amplitude values $A_{\text{REX}}^{\text{std}}\,=\,0.69\,\pm\,0.25$ are slightly higher than for the detection map obtained via the fit function $A_{\text{REX}}^{\text{fit}}\,=\,0.65\,\pm\,0.24$ in a global region of interest (ROI). The $R^2$ values are in general high and amount to $R^2\,=\,0.97$ averaged over the entire phantom. Figure 6 shows the results of the measured REX response spectra. In addition, corresponding simulation results with inclusion of relaxation were compared. For all PMF types, the measured data agree well with the simulated courses. The simulation and measurement data were normalized to their global maximum value. Clear deviations from the simulation result occur for high $f_{\text{SL}}$. Here, higher REX amplitudes were detected than predicted. Nevertheless, it becomes apparent that different PMFs can be distinguished via the acquisition of the REX response. Different PMF types as well as their variations in time-bandwidth can be identified.
\begin{figure}[htbp]
\caption*{\textbf{Figure 5)} REX detection of a Sinc stimulus. \textbf{A)} The upper row contains REX weighted images for different interaction times between the Sinc-shaped stimulus and the spin-locked magnetization. The data was adjusted for the mean signal intensity of 10 interaction timings and the real part of the complex MRI signal was presented. In \textbf{B)}, the REX signal for the two ROIs is shown as a function of interaction time (mean$\,\pm\,$std over the marked ROI). ROI\,1 is chosen to be located centrally with a $B_1^+$ greater than one. ROI\,2 is placed in the outer rim of the phantom and therefore has a comparatively higher $B_0$ deviation and $B_1^+\,<\,1$. \textbf{C)} The REX amplitude $A_{\text{REX}}^{\text{std}}$ was computed by voxel-wise calculation of the standard deviation scaled by a factor of $\sqrt{2}$. Here, the values $0.966\,\pm\,0.015$ and $0.937\,\pm\,0.016$ were obtained for the two ROIs. Alternatively, the REX amplitude $A_{\text{REX}}^{\text{fit}}$ can be calculated by conducting a voxel-wise fit with a Sine-model. Here the values $0.911\,\pm\,0.014$ and $0.888\,\pm\,0.014$ were found. For the detection, a SL frequency of 40\,Hz was selected, corresponding to the maximum spectral power density of the Bloch simulation. Other sequence parameters were: $t_{\text{SL}}\,=\,75$\,ms, $FOV\,=\,240\,\times\,240$\,mm\textsuperscript{2}, matrix: 128\,$\times$\,128.}
\includegraphics[width=0.75\textwidth]{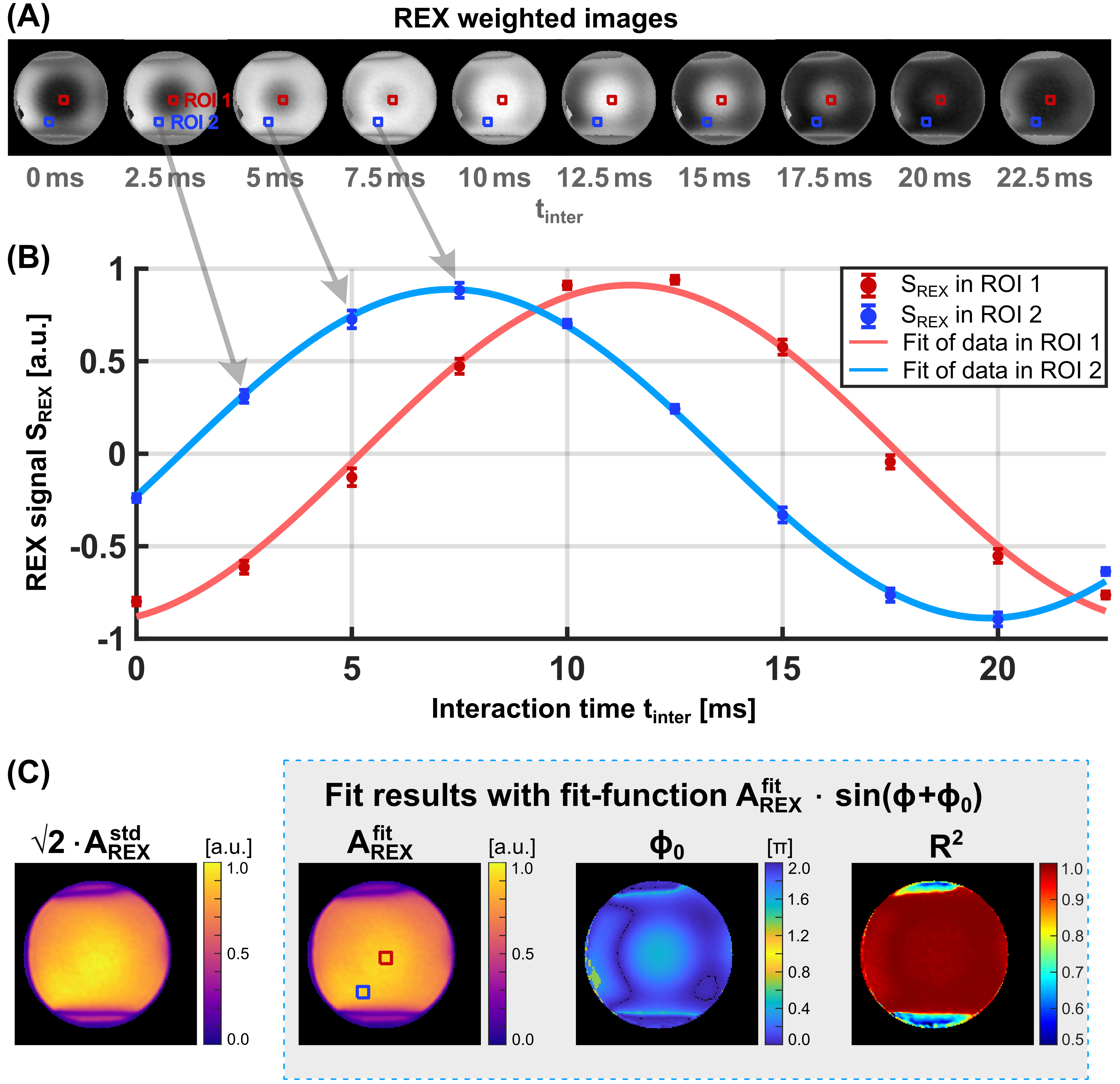}
\end{figure}
\begin{figure}[htbp]
\caption*{\textbf{Figure 6)} Comparison of simulated and measured REX response spectra for different PMF types. The simulated responses and the measured $A_{\text{REX}}$ were each normalized to their maximum value. The measured data were averaged over a ROI (depicted as mean$\,\pm\,$std) with selected field properties ($\Delta B_0\,<\,5$\,Hz, $\Delta B_1^+\,<\,5\,\%$). Here, high agreement was found between the simulated prediction of the Bloch simulation and the measured REX spectra for both, Gauss and Sinc-shaped stimuli. Different PMFs can be clearly distinguished in the experiment. The measured spectra appear to shift towards higher SL amplitudes, while the REX amplitude decreases to a constant non-zero noise level at high SL frequencies.}
\includegraphics[width=0.85\textwidth]{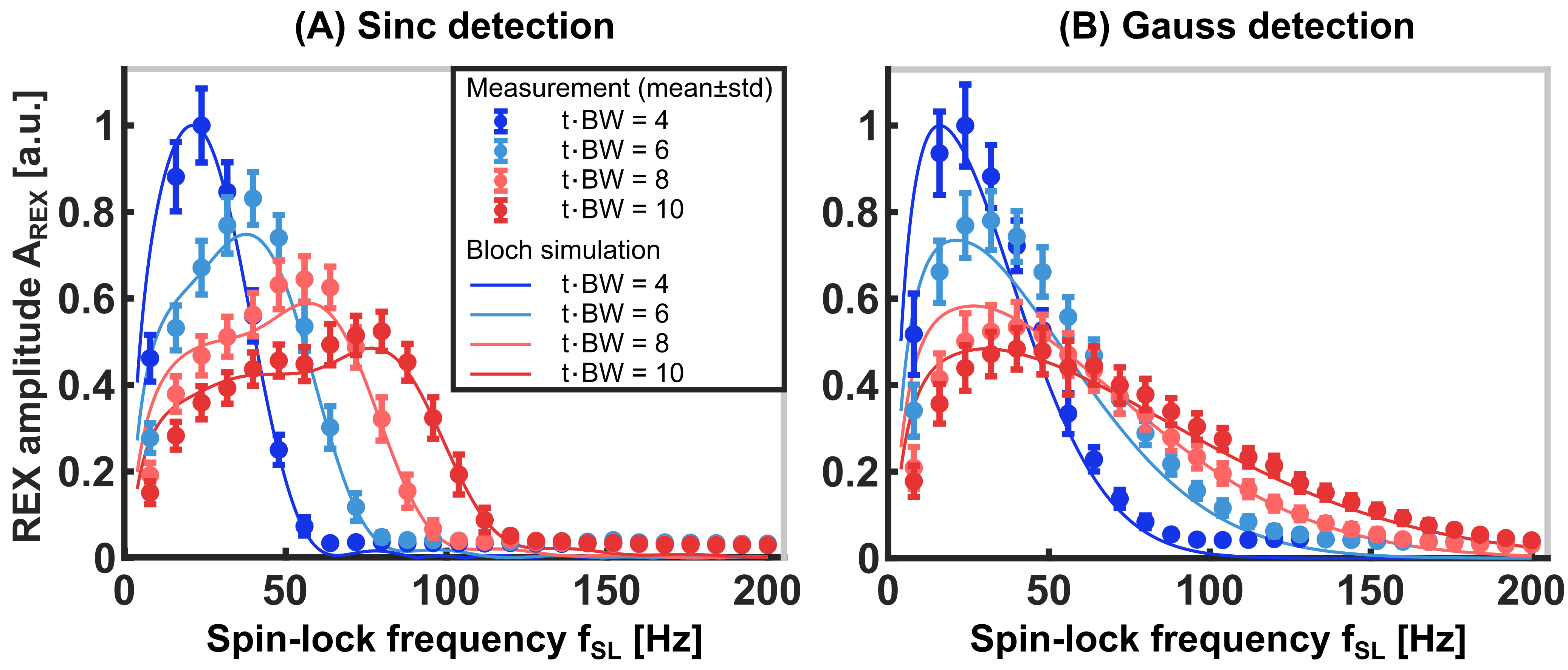}
\end{figure}

\subsection*{Detection of QRS-like fields}
The Bloch simulation results of the QRS-like stimulus are shown in Figure 7. The simulation predicts a global maximum of $A_{\text{REX}}$ for $t_{\text{SL}}\,=\,73$\,ms and $f_{\text{SL}}\,=\,12$\,Hz. These parameters were used for further experimental validation. For higher SL amplitudes, the expected REX response decreases considerably and at $100$\,Hz the expected value is reduced by a factor of $36$ compared to $12$\,Hz. Figure 8 presents the results for the detection of a PMF with $\Delta t_{\text{QRS}}\,=\,100$\,ms. The variation of the peak amplitude between $1\,\ldots\,15$\,nT shows a linear increase of $A_{\text{REX}}$ (Fig. 8B). The $R^2$ values of a voxel-wise linear fit, are close to one in the majority of the BSA phantom (Fig. 8C-D). However, $R^2$ values at the top of the phantom near the air interface were substantially reduced. The averaged $R^2$ in the marked ROI is $0.9857\,\pm\,0.0021$. Measured $A_{\text{REX}}$ values of this ROI, with field deviations of $\Delta B_0\,=\,8.62\,\pm\,0.47$\,Hz and $\Delta B_1^+\,=\,1.027\,\pm\,0.010$ were further analyzed in a box plot (Fig. 8C) and the detection results of QRS fields ($14$\,nT) with different $\Delta t_{\text{QRS}}$ along with their control experiments ($0$\,nT, $f_{\text{SL}}^{\text{off}}\,=\,100$\,Hz) were compared in Figure 9A. Significant differentiability ($p\,<\,0.001$) compared to the offresonant control can be demonstrated for all QRS durations. No significant difference was found for the $0$\,nT measurement. Figure 9B illustrates the results of the voxel-wise t-test on $p$-value heatmaps. However, REX detection yielded significantly lower statistical significance at the upper and lower areas of the phantom. In the Supporting Material (Supporting Figure 4), the $p$-values were compared with corresponding $\Delta B_0$ maps, revealing a clear correlation with offresonance effects.
\begin{figure}[htbp]
\caption*{\textbf{Figure 7)} Optimization of REX sensitivity for QRS field detection. The optimization for a stimulus with 100\,ms QRS duration was performed with parameters corresponding to myocardial tissue ($T_{1\uprho}\,=\,50$\,ms and $T_{2\uprho}\,=\,70$\,ms). The SL time and amplitude were varied between $t_{\text{SL}}\,=\,10\,\ldots\,200$\,ms and $f_{\text{SL}}\,=\,1\,\ldots\,100$\,Hz. $A_{\text{REX}}$ was calculated as the standard deviation over 20 interactions with linearly spaced interaction timings between $\pm\,40$\,ms. The optimum parameters (marked with a red cross) $t_{\text{SL}}\,=\,73$\,ms and $f_{\text{SL}}\,=\,12$\,Hz, were used for the following phantom measurements. The condition $t_{\text{SL}}\,=\,73$\,ms and $f_{\text{SL}}\,=\,100$\,Hz was used as the offresonant control. Here, the REX amplitude is reduced by a factor of 36.}
\includegraphics[width=0.65\textwidth]{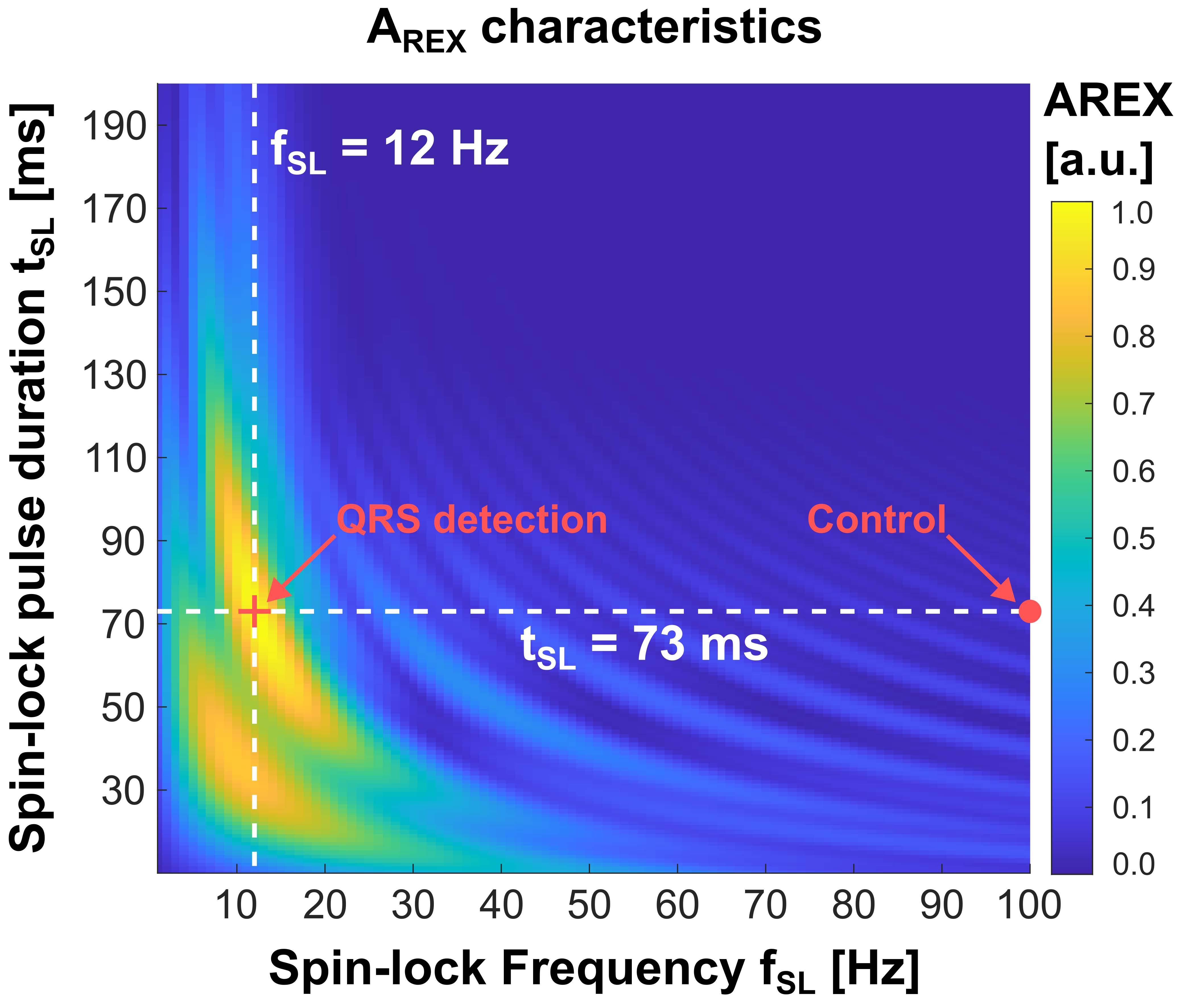}
\end{figure} 

\begin{figure}[htbp]
\caption*{\textbf{Figure 8)} Correlation of the measured REX amplitude with the QRS field strength. \textbf{A)} The experiments were performed in an axially oriented slice ($|\Delta z|\,=\,10$\,mm, 5\,mm slice thickness) in a BSA phantom placed inside a NaCl doped water-filled container. According to simulation, $f_{\text{SL}}\,=\,12$\,Hz and $t_{\text{SL}}\,=\,73$\,ms were used for the detection of a 100\,ms QRS-duration stimulus. \textbf{B)} REX amplitude maps for QRS stimulus peak amplitudes between 1\,nT and 15\,nT. The maps are calculated as the voxel-wise standard deviation over 20 REX images of varying interaction times. \textbf{C)} Map of the $R^2$ values of the voxel-wise linear fit for 15 stimulus strengths. The $R^2$ values are low in the vicinity of the water surface, since this region has high offresonances in the main magnetic field (up to $>\,80$\,Hz). The $R^2$ of the voxel-wise linear fit averaged over the ROI ($\Delta B_0\,=\,8.62\,\pm\,0.47$\,Hz, $B_1^+\,=\,1.027\,\pm\,0.010\,\%$) amounts to $0.9857\,\pm\,0.0021$. \textbf{D)} Box plot of $A_{\text{REX}}$ values in the ROI. The REX amplitude varies locally due to field properties. Yet, a linear correlation is evident.}
\includegraphics[width=0.75\textwidth]{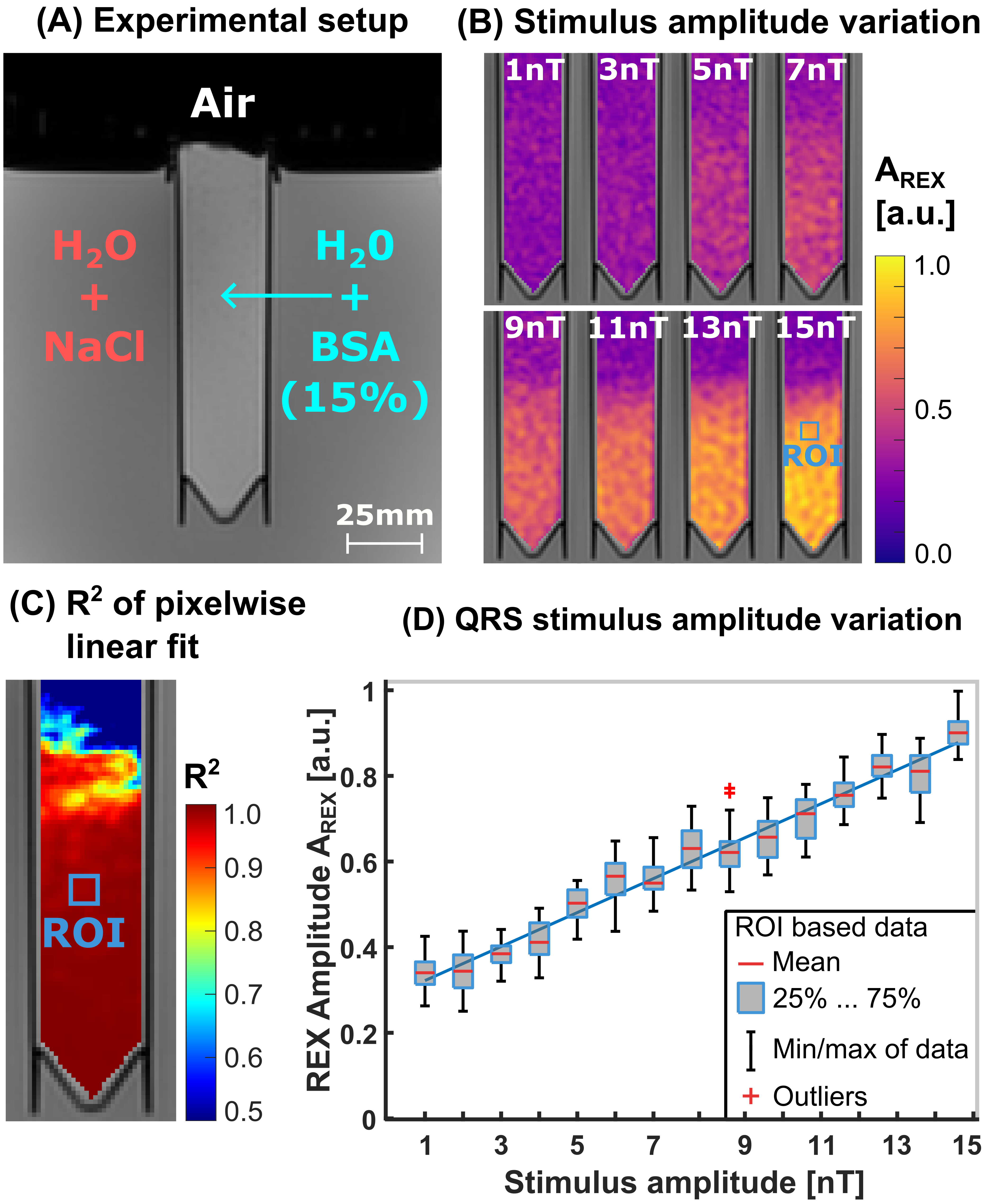}
\end{figure}
\begin{figure}[htbp]
\caption*{\textbf{Figure 9)} Statistical analysis regarding detectability of QRS-like magnetic fields. \textbf{A)} Box plot of detection experiments with stimuli of QRS durations 80$\,\ldots\,$120\,ms and two control scans. Since a detection without stimulus (0\,nT stimulus amplitude) cannot be performed in vivo, an offresonant scan needs to be used as the reference. The offresonant scan was performed with $f_{\text{SL}}\,=\,100$\,Hz and a stimulus of 100\,ms QRS duration. $A_{\text{REX}}$ was calculated based on 20 interactions. Each experiment was repeated 10 times. Data shown are taken from the ROI marked in blue. The QRS fields could all be detected significantly ($p\,<\,0.001$) compared to the control sample. The 0\,nT sample showed no significant difference. \textbf{B)} The calculated $p$-values of two-sample t-tests were depicted voxel-wise on heatmaps. The $p$-values are significantly below 5\,\% for all four stimulus durations in the majority of the phantom. Some voxels in the reference 0\,nT sample also show $p$-values below 5\,\%. The reason for the significantly higher $A_{\text{REX}}$ values in some voxels compared to the offresonant sample might be the increasing stabilizing and thus noise-reducing influence of the spin-lock pulse with higher $f_{\text{SL}}$ frequency.}
\includegraphics[width=0.75\textwidth]{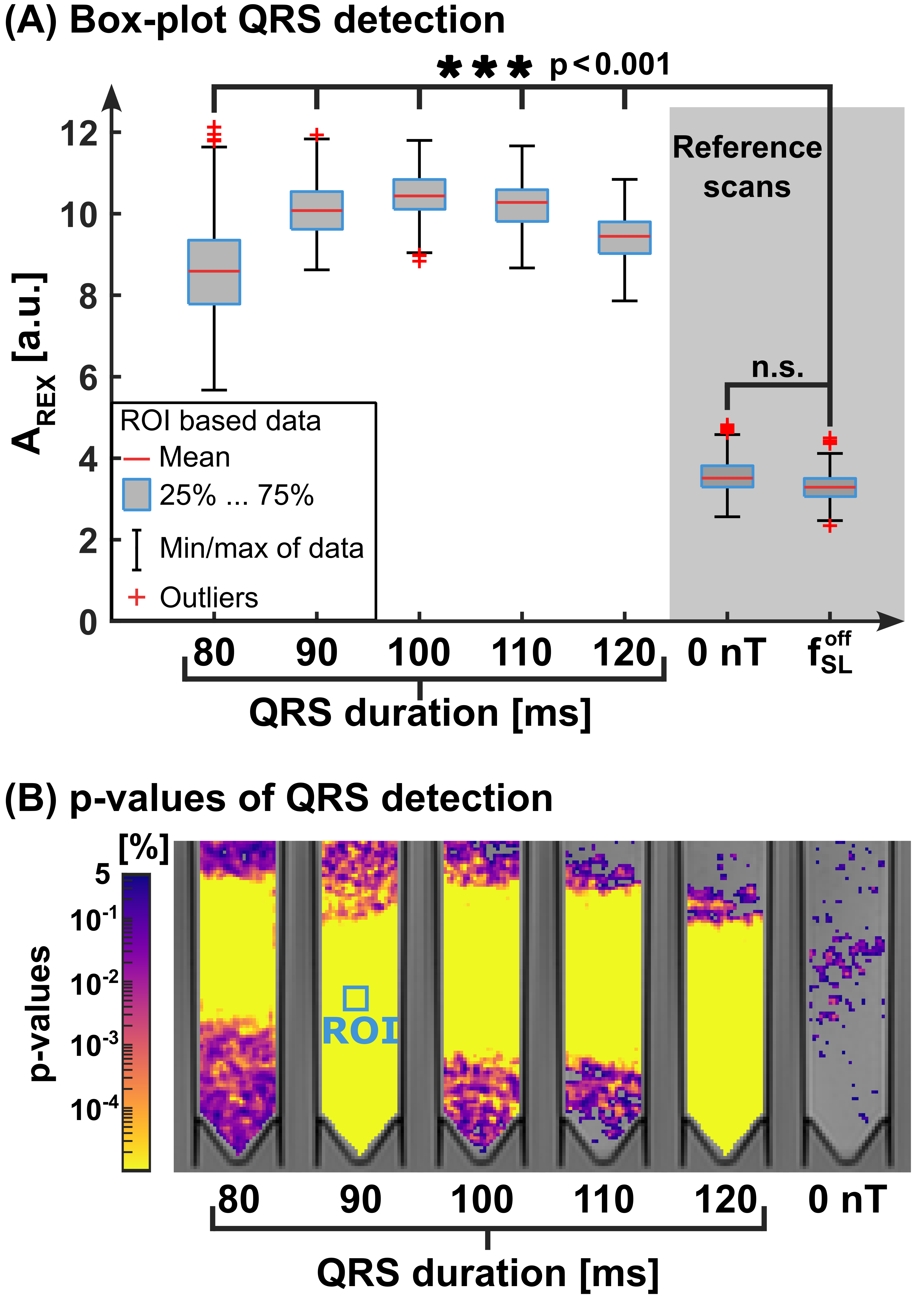}
\end{figure}
\newpage
\section*{Discussion}
The approach for the detection of sinusoidal oscillations originally published by Witzel et al. employs spin-locking to realize MRI-based imaging of ultra-weak and low-frequency oscillating magnetic fields. In the present work, this approach was combined with the concept of rotary excitation and extended as well as generalized to non-sinusoidal, pulsed magnetic fields [8, 14]. Furthermore, a concept for the direct detection of cardiac biomagnetism was presented.

\subsection*{REX postulates and limitations}
In the first step, we formulated basic postulates for generalized REX detection. We were able to validate these to a large extent in simulations and found similarities as well as differences to the interaction with sinusoidal fields. A periodic oscillation along the $z'$ axis interacts permanently with the SL component of the magnetization and continuously leads to the REX effect, generating magnetization in the $y'z'$ plane. This is mathematically equivalent to the effect of an RF pulse originating from a linearly polarized coil, where electro-magnetic radiation only occurs along a single axis [37]. Contrary to this, we demonstrated that a PMF acts like a punctual gyromagnetic force (postulate I) and can be interpreted rather as a perturbation of the SL state. However, we only studied this phenomenon in numerical simulations and an analytical analysis is still pending. A general limitation of REX detection is its sensitivity solely to the $z'$ components of magnetic fields. Still, a resonance condition can also be formulated for absorption of PMFs, but the condition is less sharply defined since the spectral characteristics are described in Fourier space by a broad spectrum of frequencies. Nevertheless, we could prove that REX detection is possible if the SL is adjusted with $f_{\text{SL}}$ matching a substantial component of the Fourier transform of the interacting field (postulate II). As observed in both simulations and measurements, relaxation effects can impair the correlation between spectral power and REX response. Particularly if long $t_{\text{SL}}$ are used, relaxation should be considered for optimizations, since the effective optimum for successful detection might be shifted to higher $f_{\text{SL}}$. A difference to the detection of pure oscillations is that the ultimately prepared REX magnetization does not depend on a relative phase between SL and stimulus but is defined by the relative timing of the interaction with the pulsed field (postulate III). This can be interpreted as a generalization of detecting sinusoidal fields at various phase relationships [13], since a phase shift can likewise be interpreted as a time difference.

\subsection*{Gauss and Sinc field detection}
Experimental validation was carried out in a phantom setup under controlled conditions. Gaussian and Sinc-shaped fields were generated with the built-in gradient system according to the tREX principle [14] (proof of principle in the Supporting Material S1). Due to the high accuracy of the gradient system, only minor deviations were observed from the nominal waveforms ($\approx\,0.5\,\%$), which were neglected for further evaluation. The relaxation times $T_{1\uprho}\,=\,221$\,ms and $T_{2\uprho}\,=\,251$\,ms of the phantom were determined with a balanced SL preparation ($t_{\text{SL}}\,=\,4...148$\,ms) [38]. The effect of dispersion was considered by calculating an average value for SL amplitudes of 10\,Hz, 50\,Hz, and 100\,Hz.\\
The evaluated detection data were able to confirm theoretical predictions to a large extent. In accordance with simulated magnetization components, highly significant variation of the measured REX contrast was observed, which primarily depends on the interaction timing. A voxel-wise signal fit with a Sine-model yielded a high level of agreement ($R^2\,>\,0.97$). This is an essential finding, since the amplitude of the oscillating REX signal is the indicator of detection. However, it was found that the $A_{\text{REX}}$ values determined by calculating the standard deviation differ from the amplitude fit. The observed discrepancy of 6.2\,\% can be attributed to the finite number of interactions. If a large number of interactions is used, this difference should be reduced. Ultimately, a trade-off must be made between accuracy and measurement time. In the specific case of ten interactions, we expect the calculated $A_{\text{REX}}$ values to be systematically overestimated by 5.4\,\%.\\
Examining the REX response spectra, there was high agreement between the measured values and simulation predictions. Due to statistical processes and the noise floor of the sequence, non-zero $A_{\text{REX}}$ was measured even without a REX effect taking place. In particular, this explains the discrepancy between measured and simulated $A_{\text{REX}}$ for frequencies with approximately zero spectral density of the stimulus. Nevertheless, clear distinction of different PMF shapes could be achieved. It is particularly noteworthy that not only the different PMF types, but even their variations regarding time-bandwidth could be distinguished. However, in order to draw conclusions from the measured REX response to the signal of the stimulus, it is not sufficient to perform an inverse Fourier transform. We expect that matching with Bloch simulations is necessary, where both sequence parameters and relaxation effects need to be addressed. For this purpose, we currently intend to develop a method for the characterization of magnetic stimuli by means of a fingerprinting approach of REX spectra using a calculated dictionary [39]. In this context, it will also be tested whether the acquisition of REX spectra can be further accelerated, since the current method still requires long measurement times ($\approx\,19$\,min).

\subsection*{QRS field detection in a phantom}
In the last part of this work, it was shown how the presented REX detection approach can be transferred to physiologically motivated PMFs such as the QRS complex observed in MCG. It is noteworthy that the optimization of the SL parameters resulted in a preparation time of only 73 ms, which means that not the entire QRS interacts with the magnetization in SL condition (Fig. 3C). This can be attributed to the consideration of relatively short relaxation times that are found in the myocardium. A tissue-like BSA phantom was used to mimic in vivo conditions. Analogous to the calibration phantom, $T_{1\uprho}\,=\,65$\,ms and $T_{2\uprho}\,=\,90$\,ms were determined in a pre-experiment. The $T_{1\uprho}$ value is thus slightly higher than in the myocardium at 3\,T and rather resembles the expected range at 1.5\,T [34]. A major difference to the Gauss and Sinc detection was that the interaction times were not varied linearly, but via the random RR variation from an ECG data set. The developed sequence is thus already suitable for in vivo application with a prospective trigger. It was found that the natural RR variation is sufficient to shift the interaction timings for detection. With the current sequence design and the concept of t-test validation, a total of 400 REX-weighted images are required for 10 detection experiments and 10 controls (20 interactions each). Given an average RR of 850\,ms and a rejection rate of 45\,\%, the time required for an examination would amount to $\approx\,20$\,min. Here it has already been considered that only every second heartbeat can be used due to prospective triggering.\\
With the REX data, significant detections for QRS durations of $80\,\ldots\,$120\,ms ($p\,<\,0.001$) and linearity of the measured $A_{\text{REX}}$ values for the range $1\,\ldots\,15$\,nT ($R^2\,>\,0.98$) were confirmed. However, less reliable results were obtained at the upper and lower ends of the phantom. As can be seen in the Supporting Material (Supporting Fig. 4), stronger $B_0$ inhomogeneities were present at these areas. The reduction of the $R^2$ values as well as the increase of the $p$-values coincide closely with the offresonances found in the field maps. This emphasizes the importance of a good shim for reliable detectability of PMFs. According to the simulation of QRS detection for different offresonances (Supporting Fig. 5), the FWHM of $A_{\text{REX}}$ is 23.8\,Hz. In addition to this dependency on $\Delta B_0$, it was previously noticed in the calibration phantom that the $A_{\text{REX}}$ values indicated a radially symmetric distribution. This suggests an effect of central brightening typical for $B_1^+$ field inhomogeneities at 3\,T [40]. The susceptibility of SL based field detection to static field inhomogeneities has already been discussed in literature [14, 41]. This issue is a major challenge and needs to be addressed in future studies to enable applications involving in vivo PMF detection. In the present work, only a simple SL preparation consisting of a single continuous wave pulse was used, which is known to be highly susceptible to static field imperfections [42]. In order to achieve sufficient robustness, adopting compensation ideas developed for $T_{1\uprho}$ quantification might be feasible. Future studies will need to investigate the detection capability of self-compensated SL preparations such as rotary-echo, composite, or balanced spin-locking [38, 43, 44].

\subsection*{Towards direct imaging of cardiac conduction}
The establishment of a method for non-invasive direct detection of cardiac conduction with high spatial resolution would be a great benefit for basic research and clinical diagnostics. The spatial resolution of routinely applied catheter examination is limited since the point-by-point scanning of the endocardium causes lengthy screenings [45]. In addition, detection is limited to potentials that are tapped at the surface of the myocardium and limited conclusions can currently be drawn about processes inside the myocardial tissue. In contrast, MCG can be used for non-invasive cardiac field detection. For this purpose, highly sensitive SQUIDs (superconducting quantum interference devices) or OPMs (optically pumped magnetometers) are used, and magnetically shielded labs are required, since the detected fields are only in the pT range [5, 25, 46]. However, the specific MCG hardware is rarely available in clinical environments and in addition, there is one more crucial disadvantage associated with MCG. Due to the inverse problem, spatial resolution is strongly limited, and the magnetometers perform field detection from a certain distance [1]. In a study on an isolated guinea pig heart, a field of $\approx\,10$\,pT was observed from a distance of 15\,mm and a field of $\approx\,100$\,pT was observed at 5\,mm [25]. Another recent study shows field detection in an open chest setup on rats [47]. The fields were measured in close proximity to the heart and the peak magnitudes reached values of 20\,nT.\\
The major benefit of REX-based field detection is that, on the one hand, the spatial resolution of MRI can be utilized, and, on the other hand, the magnetic fields can be sensed non-invasively at the origin of their source. Therefore, achieving the high sensitivity of magnetometers is not necessary since the decay with distance does not occur. In this context, our results are promising, as detections in the lower nT range have already been demonstrated. Currently, we identify two key points for the further development of a method of in vivo REX detection in the heart. The SL preparation needs to be optimized regarding the spectral properties of the cardiac fields actually present at the origin. Detection with solid state quantum sensors suggest the QRS shapes have shorter durations measured in close proximity than at the chest [47]. Since a shortened QRS duration results in spectral signal components of higher frequency, higher $f_{\text{SL}}$ would have to be employed. As the influence of static offresonances decreases with $f_{\text{SL}}$, this might lead to more favorable experimental conditions. The second point is, that significant optimizations are still needed to transfer the concept to the demanding in vivo setup. Our phantom experiments were carried out under optimal conditions, with no disruptive factors such as cardiac motion, blood flow or breathing. We therefore expect that the development of compensation techniques will be necessary and utmost care must be taken regarding shim adjustments. Therefore, we cannot yet definitively confirm that the sensitivity of REX to cardiac fields is given. A meaningful intermediate step would be the detection of cardiac fields in the isolated heart [25]. The optimal experimental setup involves an ex vivo heart with inhibited muscle contraction but physiological electrical activity [48]. Furthermore, we intend to carry out initial ex vivo and in vivo trials on devices with 0.55\,T and 1.5\,T, as lower field deviations are expected for both $B_0$ and $B_1^+$ [49, 50].\\\,\\

\newpage
\section*{Conclusion}
In this work, a novel concept of MRI based imaging of ultra-weak and low-frequency pulsed magnetic fields was introduced. The practical feasibility of spatially resolved detection of PMFs with magnitudes in the nT range was demonstrated and high agreement with results from Bloch simulations was found. The acquisition of REX response spectra, performed via variation of the SL amplitude, allows a characterization of magnetic fluctuations regarding their spectral properties. Besides Gauss- and Sinc-shaped fields, QRS-like fields could be detected. Our work indicates that REX potentially enables direct imaging of cardiac fields and thus may provide a useful tool for non-invasive assessment of cardiac conduction and, ultimately, rhythm disorders. REX-based detection thus offers fascinating opportunities for clinical applications and basic medical research.

\section*{Acknowledgements}
We would like to thank the EZRT/MRB team of the Fraunhofer Institute for the technical support on the MRI system and the developers of the open source Pulseq framework, which enabled efficient prototyping in this project. We would further like to thank the manuscript reviewers for their valuable feedback, which significantly progressed the presented method of QRS detection.

\section*{Data availability statement}
The datasets used and/or analyzed during the current study are available from the corresponding author on reasonable request. An exemplary Pulseq source code for the detection of magnetic peak-like fields will be provided in an open-access Github repository at the time of acceptance of this manuscript.

\section*{Conflict of Interest}
The authors declare that they have no competing interests to disclose.

\section*{Authors' contributions}
P. Albertova and M. Gram contributed equally to this work; guarantors of integrity of entire study: P. Nordbeck, PM. Jakob; study concepts and study design: all authors; data acquisition: M. Gram, P. Albertova; data analysis: M. Gram, P. Albertova; manuscript drafting and manuscript editing: all authors.

\newpage
\section*{References}
\textbf{[1]} Williamson SJ, KaufmanL, Biomagnetism. J Magn Magn Mater. 1981;22(2):129-201.\\
doi:10.1016/0304-8853(81)90078-0
\\\textbf{[2]} Wheless JW, Castillo E, Maggio V, et al. Magnetoencephalography (MEG) and magnetic source imaging (MSI). Neurologist. 2004;10(3):138-153. doi: 10.1097/01.nrl.0000126589.21840.a1
\\\textbf{[3]} Hämäläinen M, Hari R, Ilmoniemi RJ, Knuutila J, Lounasmaa OV. Magnetoencephalography-theory, instrumentation, and applications to noninvasive studies of the working human brain. Rev Mod Phys. 1993;65(2):413-497. doi: 10.1103/RevModPhys.65.413
\\\textbf{[4]} Baule G, McFee R. Detection of the magnetic field of the heart. Am Heart J. 1963;66:95-96. doi:10.1016/0002-8703(63)90075-9
\\\textbf{[5]} Nakaya Y, Mori H. Magnetocardiography. Cli. Phys Physiol Meas. 1992;13(3):191.\\
doi: 10.1088/0143-0815/13/3/001
\\\textbf{[6]} Bandettini PA, Petridou N, Bodurka J. Direct detection of neuronal activity with MRI: Fantasy, possibility, or reality?. Appl Magn Reson. 2005. 29, 65–88. doi: 10.1007/BF03166956
\\\textbf{[7]} Kraus RH Jr, Volegov P, Matlachov A, Espy M. Toward direct neural current imaging by resonant mechanisms at ultra-low field. Neuroimage. 2008 Jan 1;39(1):310-317.\\
doi: 10.1016/j.neuroimage.2007.07.058
\\\textbf{[8]} Witzel T, Lin FH, Rosen BR, Wald LL. Stimulus-induced Rotary Saturation (SIRS): a potential method for the detection of neuronal currents with MRI. Neuroimage. 2008 Oct 1;42(4):1357-65. doi: 10.1016/j.neuroimage.2008.05.010
\\\textbf{[9]} Redfield AG. Nuclear magnetic resonance saturation and rotary saturation in solids. Physical Review. 1955. 98:1787-1809. https://doi.org/10.1103/PhysRev.98.1787
\\\textbf{[10]} Gilani IA, Sepponen R. Quantitative rotating frame relaxometry methods in MRI. NMR Biomed. 2016 Jun;29(6):841-61. doi: 10.1002/nbm.3518
\\\textbf{[11]} Jiang X, Sheng J, Li H, et al. Detection of subnanotesla oscillatory magnetic fields using MRI. Magn Reson Med. 2016;75(2):519-526. doi:10.1002/mrm.25553
\\\textbf{[12]} Chai Y, Bi G, Wang L, et al. Direct detection of optogenetically evoked oscillatory neuronal electrical activity in rats using SLOE sequence. Neuroimage. 2016;125:533-543.\\
doi:10.1016/j.neuroimage.2015.10.058
\\\textbf{[13]} Truong TK, Roberts KC, Woldorff MG, Song AW. Toward direct MRI of neuro-electro-magnetic oscillations in the human brain. Magn Reson Med. 2019;81(6):3462-3475. doi:10.1002/mrm.27654
\\\textbf{[14]} Gram M, Albertova P, Schirmer V, et al. Towards robust in vivo quantification of oscillating biomagnetic fields using Rotary Excitation based MRI. Sci Rep. 2022;12(1):15375. Published 2022 Sep 13. doi:10.1038/s41598-022-19275-5
\\\textbf{[15]} Ueda H, Ito Y, Oida T, Taniguchi Y, Kobayashi T. Detection of tiny oscillatory magnetic fields using low-field MRI: A combined phantom and simulation study. J Magn Reson. 2020;319:106828. doi:10.1016/j.jmr.2020.106828
\\\textbf{[16]} Lopes da Silva F. Neural mechanisms underlying brain waves: from neural membranes to networks. Electroencephalogr Clin Neurophysiol. 1991;79(2):81-93. doi:10.1016/0013-4694(91)90044-5
\\\textbf{[17]} Nagahara S, Ueno M, Kobayashi T, Spin-Lock Imaging for Direct Detection of Oscillating Magnetic Fields with MRI: Simulations and Phantom Studies. Adv. Biomed. Eng. 2013, 2:63-71. doi:10.14326/abe.2.63
\\\textbf{[18]} Roth BJ, Wikswo JP Jr. The magnetic field of a single axon. A comparison of theory and experiment. Biophys J. 1985;48(1):93-109. doi:10.1016/S0006-3495(85)83763-2
\\\textbf{[19]} Wikswo JP, Barach JP, Freeman JA. Magnetic field of a nerve impulse: first measurements. Science. 1980;208(4439):53-55. doi:10.1126/science.7361105
\\\textbf{[20]} Okada YC, Lauritzen M, Nicholson C. Magnetic field associated with neural activities in an isolated cerebellum. Brain Res. 1987;412(1):151-155. doi:10.1016/0006-8993(87)91451-x
\\\textbf{[21]} Hart G. Biomagnetometry: imaging the heart's magnetic field. Br Heart J. 1991;65(2):61-62. doi:10.1136/hrt.65.2.61
\\\textbf{[22]} Mäntynen V, Konttila T, Stenroos M. Investigations of sensitivity and resolution of ECG and MCG in a realistically shaped thorax model. Phys Med Biol. 2014;59(23):7141-7158. doi:10.1088/0031-9155/59/23/7141
\\\textbf{[23]} Price A, Santucci P. Electrophysiology procedures: weighing the factors affecting choice of anesthesia. Semin Cardiothorac Vasc Anesth. 2013;17(3):203-211. doi:10.1177/1089253213494023
\\\textbf{[24]} Majeed H, Sattar Y. Electrophysiologic Study Indications and Evaluation. In: StatPearls. Treasure Island (FL): StatPearls Publishing; April 3, 2023.
\\\textbf{[25]} Jensen K, Skarsfeldt MA, Stærkind H, et al. Magnetocardiography on an isolated animal heart with a room-temperature optically pumped magnetometer. Sci Rep. 2018;8(1):16218. Published 2018 Nov 1. doi:10.1038/s41598-018-34535-z
\\\textbf{[26]} Ueda H, Seki H, Ito Y, Oida T, Taniguchi Y, Kobayashi T. Dynamics of magnetization under stimulus-induced rotary saturation sequence. J Magn Reson. 2018;295:38-44.\\
doi:10.1016/j.jmr.2018.07.004
\\\textbf{[27]} Layton KJ, Kroboth S, Jia F, et al. Pulseq: A rapid and hardware-independent pulse sequence prototyping framework. Magn Reson Med. 2017;77(4):1544-1552. doi:10.1002/mrm.26235
\\\textbf{[28]} Pauly J, Le Roux P, Nishimura D, Macovski A. Parameter relations for the Shinnar-Le Roux selective excitation pulse design algorithm [NMR imaging]. IEEE Trans Med Imaging. 1991;10(1):53-65. doi: 10.1109/42.75611
\\\textbf{[29]} Schuenke P, Koehler C, Korzowski A, et al. Adiabatically prepared spin-lock approach for T1$\rho$-based dynamic glucose enhanced MRI at ultrahigh fields. Magn Reson Med. 2017;78(1):215-225. doi:10.1002/mrm.26370
\\\textbf{[30]} Schuenke P, Windschuh J, Roeloffs V, Ladd ME, Bachert P, Zaiss M. Simultaneous mapping of water shift and B1 (WASABI)-Application to field-Inhomogeneity correction of CEST MRI data. Magn Reson Med. 2017 Feb;77(2):571-580. doi: 10.1002/mrm.26133
\\\textbf{[31]} Sniderman A. Atherosclerosis Risk Factors. James J. Maciejko. Clin Chem. 2004; 51(8): 1568-1568. doi: 10.1373/clinchem.2004.040527
\\\textbf{[32]} Maury P, Lematte E, Derval N, et al. Prevalence and long-term prognosis of patients with 'narrower than normal' QRS complexes. Europace. 2018;20(4):692-697. doi:10.1093/europace/euw401
\\\textbf{[33]} Xu D, Roth BJ. The Magnetic Field Produced by the Heart and Its Influence on MRI. Mathematical Problems in Engineering. 2017; 2017:1–9. doi: 10.1155/2017/3035479
\\\textbf{[34]} Bustin A, Witschey WRT, van Heeswijk RB, Cochet H, Stuber M. Magnetic resonance myocardial T1$\rho$ mapping : Technical overview, challenges, emerging developments, and clinical applications. J Cardiovasc Magn Reson. 2023 Jun 19;25(1):34. doi: 10.1186/s12968-023-00940-1
\\\textbf{[35]} Gram M, Albertova P, Gutjahr FT, Jakob PM, Bauer WR, Nordbeck P, Christa M. Myocardial T2$\rho$ Mapping in Small Animals: Comparison of Balanced Spin-Lock and Malcolm-Levitt Preparations. In Proceedings of the Annual Meeting of the ISMRM. Toronto. 2023. 0173
\\\textbf{[36]} Morales S, Corsi MC, Fourcault W, et al. Magnetocardiography measurements with 4He vector optically pumped magnetometers at room temperature. Phys Med Biol. 2017;62(18):7267-7279. Published 2017 Aug 21. doi:10.1088/1361-6560/aa6459
\\\textbf{[37]} Glover GH, Hayes CE, Pelc NJ, Edelstein WA, Mueller OM, Hart HR, Hardy CI, O'Donnell M, Barber WD. Comparison of linear and circular polarization for magnetic resonance imaging. J Magn Reson. 1985;64:255-270. doi: 10.1016/0022-2364(85)90349-X
\\\textbf{[38]} Gram M, Seethaler M, Gensler D, Oberberger J, Jakob PM, Nordbeck P. Balanced spin-lock preparation for B1 -insensitive and B0 -insensitive quantification of the rotating frame relaxation time T1$\rho$. Magn Reson Med. 2021 May;85(5):2771-2780. doi: 10.1002/mrm.28585
\\\textbf{[39]} Ma D, Gulani V, Seiberlich N, Liu K, Sunshine JL, Duerk JL, Griswold MA. Magnetic resonance fingerprinting. Nature. 2013 Mar 14;495(7440):187-92. doi: 10.1038/nature11971
\\\textbf{[40]} Bernstein MA, Huston J 3rd, Ward HA. Imaging artifacts at 3.0T. J Magn Reson Imaging. 2006 Oct;24(4):735-46. doi: 10.1002/jmri.20698
\\\textbf{[41]} Coletti C, Domsch S, Vos F, Weingärtner S. Functional MRI of neuro-electro-magnetic oscillations: Statistical processing in the presence of system imperfections. IEEE-EMBS Conf Biomed Eng Sci. 2021. 172–177. doi: 10.1109/IECBES48179.2021.9398751
\\\textbf{[42]} Chen W. Errors in quantitative T1rho imaging and the correction methods. Quant Imaging Med Surg. 2015 Aug;5(4):583-91. doi: 10.3978/j.issn.2223-4292.2015.08.05
\\\textbf{[43]} Charagundla SR, Borthakur A, Leigh JS, Reddy R. Artifacts in T(1rho)-weighted imaging: correction with a self-compensating spin-locking pulse. J Magn Reson. 2003 May;162(1):113-21. doi: 10.1016/s1090-7807(02)00197-0
\\\textbf{[44]} Witschey WR, Borthakur A, Elliott MA, et al. Compensation for spin-lock artifacts using an off-resonance rotary echo in T1rhooff-weighted imaging. Magn Reson Med. 2007;57(1):2-7.\\
doi:10.1002/mrm.21134
\\\textbf{[45]} Steinberg J, Jais P, Calkins H. Practical Guide to Catheter Ablation of Atrial Fibrillation. Wiley 201. isbn: 9781118658505
\\\textbf{[46]} Alem O, Sander TH, Mhaskar R, et al. Fetal magnetocardiography measurements with an array of microfabricated optically pumped magnetometers. Phys Med Biol. 2015;60(12):4797-4811. doi:10.1088/0031-9155/60/12/4797
\\\textbf{[47]} Arai K, Kuwahata A, Nishitani D, et al. Millimetre-scale magnetocardiography of living rats with thoracotomy. Commun Phys. 2022;5:200. doi: 10.1038/s42005-022-00978-0
\\\textbf{[48]} Vaillant F, Magat J, Bour P, et al. Magnetic resonance-compatible model of isolated working heart from large animal for multimodal assessment of cardiac function, electrophysiology, and metabolism. Am J Physiol Heart Circ Physiol. 2016;310(10):H1371-H1380. doi:10.1152/ajpheart.00825.2015
\\\textbf{[49]} Hock M, Terekhov M, Stefanescu MR, et al. B0 shimming of the human heart at 7T. Magn Reson Med. 2021;85(1):182-196. doi:10.1002/mrm.28423
\\\textbf{[50]} Reeder SB, Faranesh AZ, Boxerman JL, McVeigh ER. In vivo measurement of T*2 and field inhomogeneity maps in the human heart at 1.5 T. Magn Reson Med. 1998;39(6):988-998.\\
doi:10.1002/mrm.1910390617

\newpage
\section*{Supplementary Material}
\subsection*{S.1 Validation of the gradient waveforms for the tREX application}
The stimulus fields of the detection experiments were applied with the z-gradient in an offcenter slice. This approach allows the investigation of reproducible stimuli with different waveforms and adjustable amplitudes. The advantage compared to the experimental setup of a phantom experiment with a loop coil or dipole antenna is that no additional hardware is required for REX measurements, no external noise sources are introduced into the scanner and the gradient system can be controlled very precisely. For our experiments, gradient waveforms were used to obtain ultra-low fields in the range of $1\,\ldots\,100$\,nT in a 10\,mm offcenter slice. Thus, the gradients had to be driven at their lower limit and produced waveforms of $0.1\,\ldots\,10$\,nT/mm. For control these waveforms were measured via the observed phase evolution in a thin slice far away from the isocenter. This concept originates from the calibration of k-space trajectories according to Duyn et al. [S1]. Supplementary Fig. 1 shows the waveforms measured in a 2\,mm thick slice at a distance of 50\,mm from the isocenter. The measurements were averaged 100 times and the phase evolution was smoothed by means of a moving average filter. It could be verified that the magnetic tREX stimuli show high accuracy and the produced peak magnitudes deviate only 0.51\,\% from the nominal values (Supplementary Fig. 1).

\begin{figure}[H]
\caption*{\textbf{Supplementary Figure 1)} Measured gradient trajectories for a Sinc- and Gauss-shaped stimulus of 50\,ms duration at $t\,\cdot\, BW\,=\,6$. The validation was performed on a spherical phantom (diameter 17.5\,cm) of demineralized water doped with 1.25\,g/L NiSO$_4$ in a slice with 50\,mm offcenter distance and 2\,mm slice thickness. The gradient amplitude was varied between 0.1 and 10\,nT/mm which corresponds to a peak amplitude of 1 to 100\,nT in a slice at 10\,mm offcenter distance.}
\includegraphics[width=0.98\textwidth]{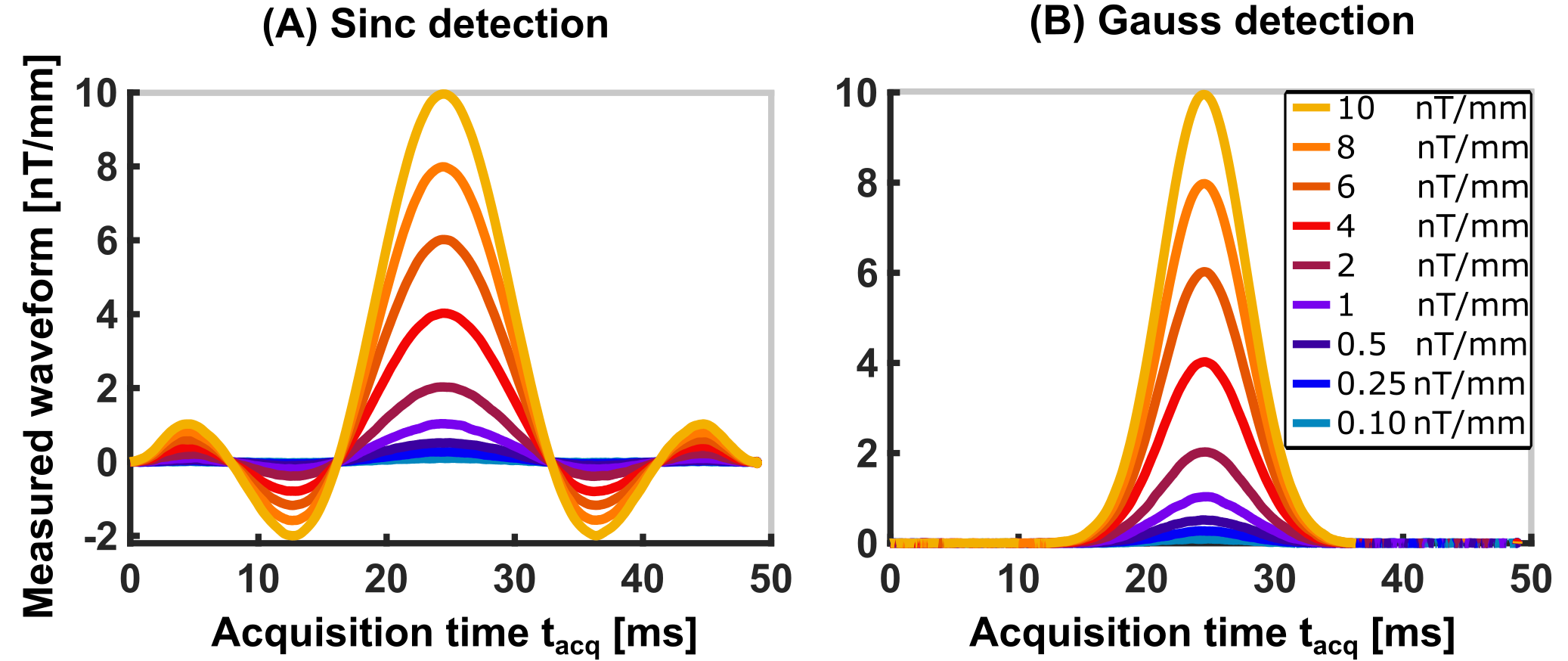}
\end{figure}

\begin{figure}[H]
\caption*{\textbf{Supplementary Figure 2)} Control of the peak magnitudes of the calibrated waveforms. The observed values were compared with the nominal values. An average deviation of 0.51\,\% was determined.}
\includegraphics[width=0.75\textwidth]{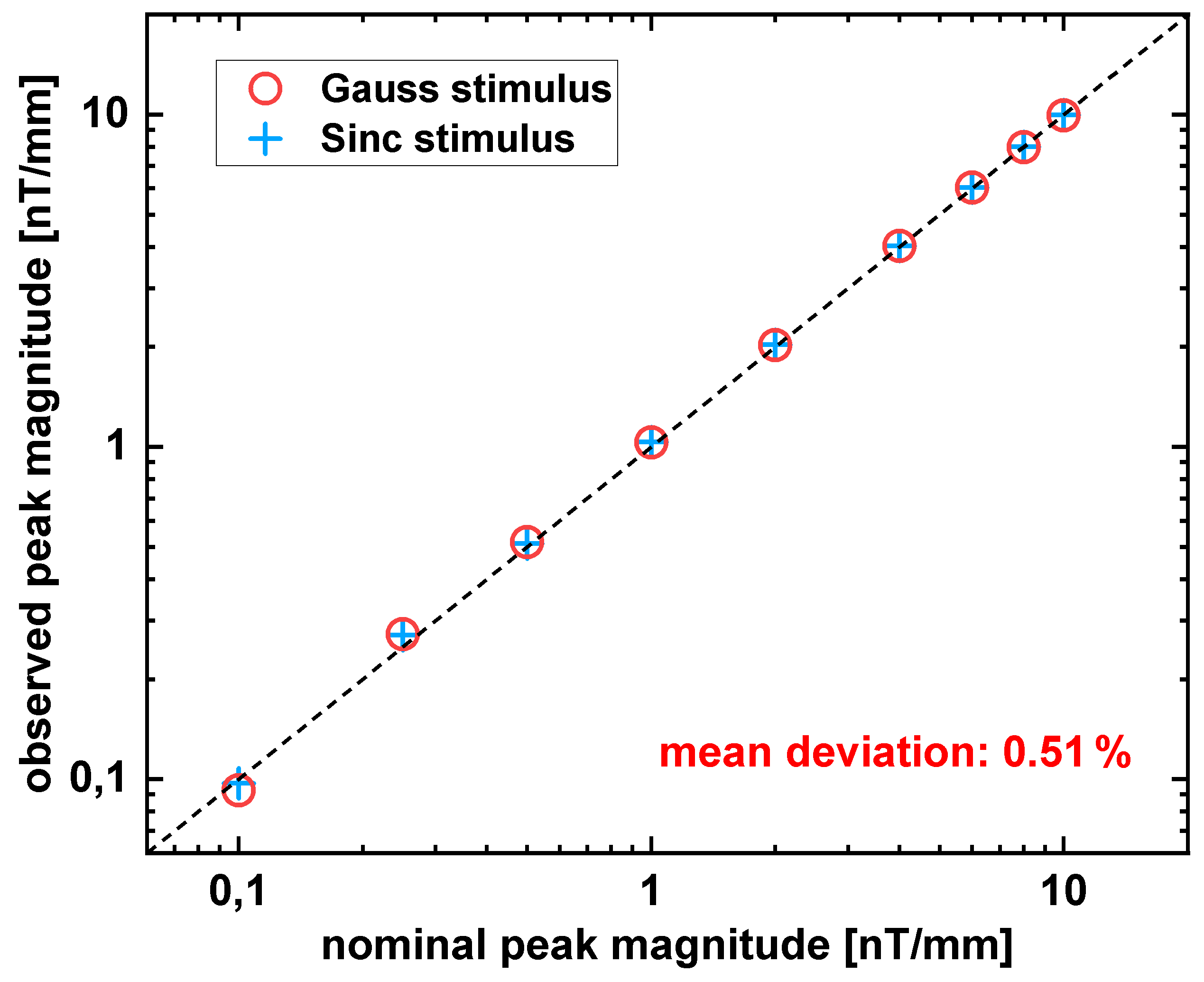}
\end{figure}

\newpage
\subsection*{S.2 Extended measurement results: $B_0$ and $B_1^+$ field inhomogeneities}
Since the influence of field inhomogeneities on REX-based field detection is severe, knowledge of the field properties is important for the interpretation of measured $A_{\text{REX}}$ maps. This section provides the field maps corresponding to the experiments in Fig. 9 of the main manuscript. 
Field maps were recorded using a WASABI sequence with offresonances between -2 to 2\,ppm in steps of 0.05\,ppm [S2]. The experiments in Fig. 9 of the main paper were performed without moving the phantom, but a new shim and frequency adjustment was performed between the lengthy sequence blocks (different QRS durations, the reference measurement without stimulus or with offresonant SL pulse) to reduce errors caused by field drift. The $B_1^+$ map shown in Supplementary Fig. 3 is relevant for all experiments of QRS detection. It reports values corrected by a global factor of 1.23, as the set SL frequency was adjusted by this value based on a $B_1^+$ field analysis prior to the experiments.
In Supplementary Fig. 4, all corresponding $B_0$ maps of the different detection experiments are shown.  Areas of high offresonance largely correspond to regions in which the t-test confirms detectability only to a limited extent or where even no statistically significant distinction from the reference measurement is possible.

\begin{figure}[H]
\caption*{\textbf{Supplementary Figure 3)} Corrected transmission field amplitude in the BSA phantom. Due to the correction with a factor of 1.23 determined from preliminary experiments and the installation of the small sample in a large water phantom containing NaCl, the transmitted SL field deviates only slightly from its nominal value.}
\includegraphics[width=0.35\textwidth]{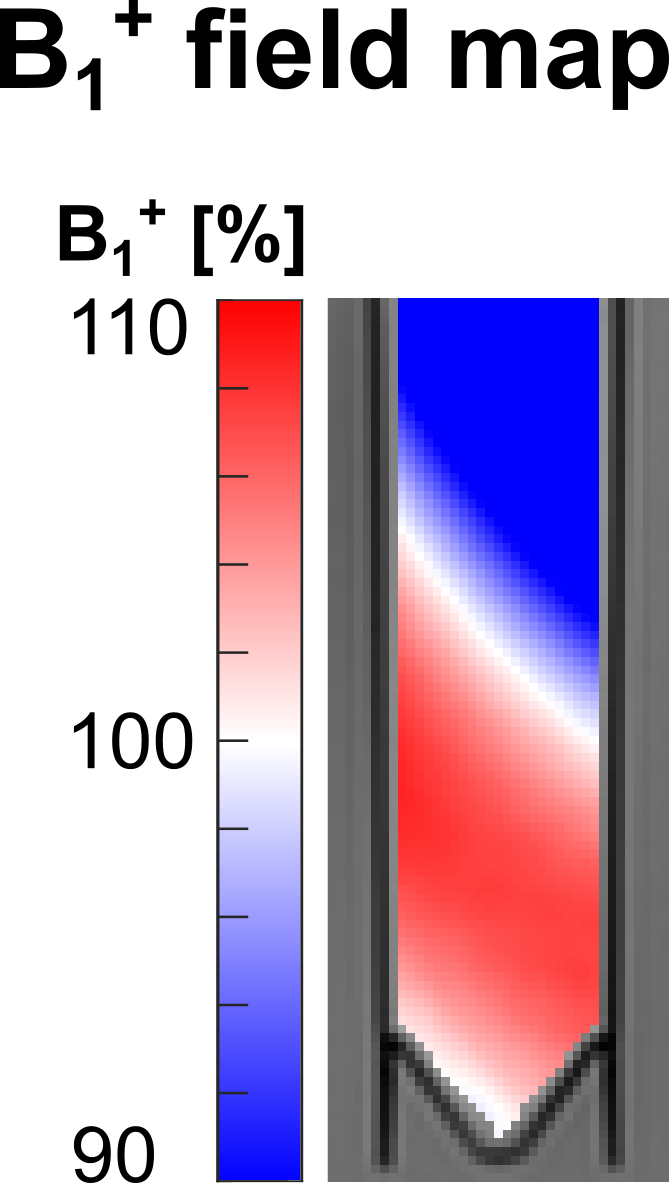}
\end{figure}

\begin{figure}[H]
\caption*{\textbf{Supplementary Figure 4)} Maps of the deviation in the main magnetic field strength, supplementing Fig. 9 of the main paper. For clarity, the $p$-value maps from Fig. 9 are shown. The errors from the offresonant measurement are propagated to all $p$-value maps, as the offresonant measurement is used as a reference for the hypothesis test. The correspondence of the areas of high offresonance with regions of lower statistical significance indicated by elevated $p$-values is apparent. This emphasizes the need for good field properties for REX based field detection.}
\includegraphics[width=0.98\textwidth]{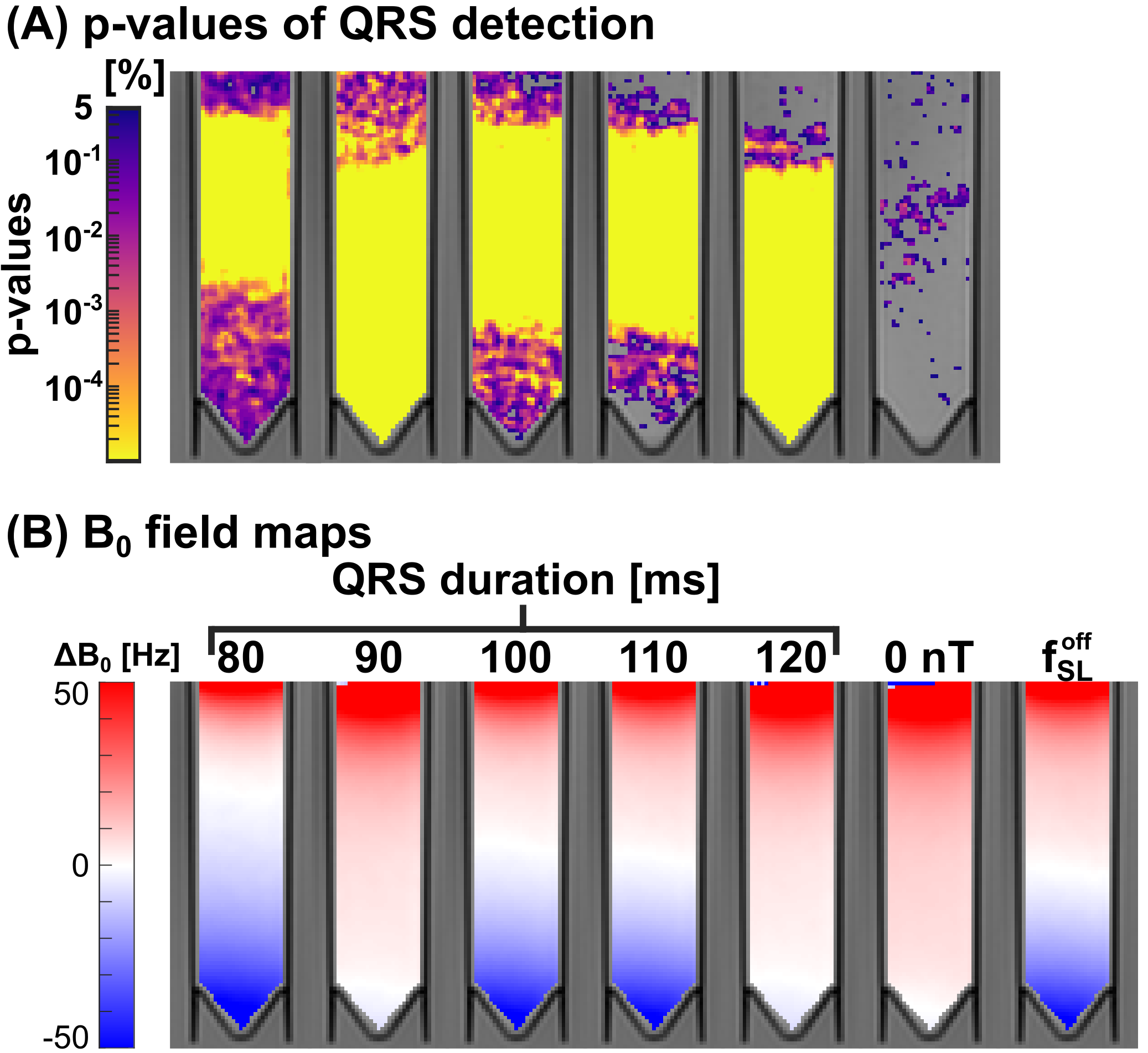}
\end{figure}

\newpage
\subsection*{S.3 Bloch simulation: influence of $B_0$ field inhomogeneities for QRS detection}
Spin-lock based field detection requires a high homogeneity of the main magnetic field in the probe volume, since both deviations in the main magnetic field and the transmitting field disrupt REX-based field detection. Deviations from the nominal transmission field amplitude change the resonant frequency defined by the SL pulse. Static offresonances, on the other hand, lead to an effective tilting of the SL axis. This tilt causes a non-vanishing longitudinal magnetization component in the absence of a stimulus. 
The results of a Bloch simulation for offresonances in the range -100 to 100\,Hz show the REX amplitude for a field detection of a QRS-like stimulus with 14\,nT magnitude. Assumed parameters including relaxation times ($T_{1\uprho}$\,=\,50\,ms, $T_{2\uprho}$\,=\,70\,ms) are chosen in view of the in vivo experiment. The resulting low FWHM of 23.8\,Hz illustrates the importance of good shimming and the necessity of investigating SL preparation modules compensated for field inhomogeneities [S3, S4, S5]. For future in vivo application, it might be advisable to use scanners with field strengths of 0.55\,T or 1.5\,T, since the offresonances decrease linearly with the field strength [S6]. At such field strengths, realistic $\Delta B_0$ values in the heart could be 10\,-\,30\,Hz [S7, S8, S9], assuming linear decrease with field strength.

\begin{figure}[H]
\caption*{\textbf{Supplementary Figure 5)} Simulated effect of the presence of static $B_0$ inhomogeneities on the detection of QRS-like magnetic fields. Detection parameters of spin-lock frequency 12\,Hz and spin-lock pulse duration 73\,ms were chosen according to simulative optimization for maximum REX amplitude (Fig. 7, main paper).  The interaction timing of the QRS-like stimulus with an amplitude of 14\,nT was linearly varied in 30 steps. The REX signal has a Full-Width-Half-Maximum of 23.8\,Hz with regard to deviations of the main magnetic field from its nominal strength.}
\includegraphics[width=0.75\textwidth]{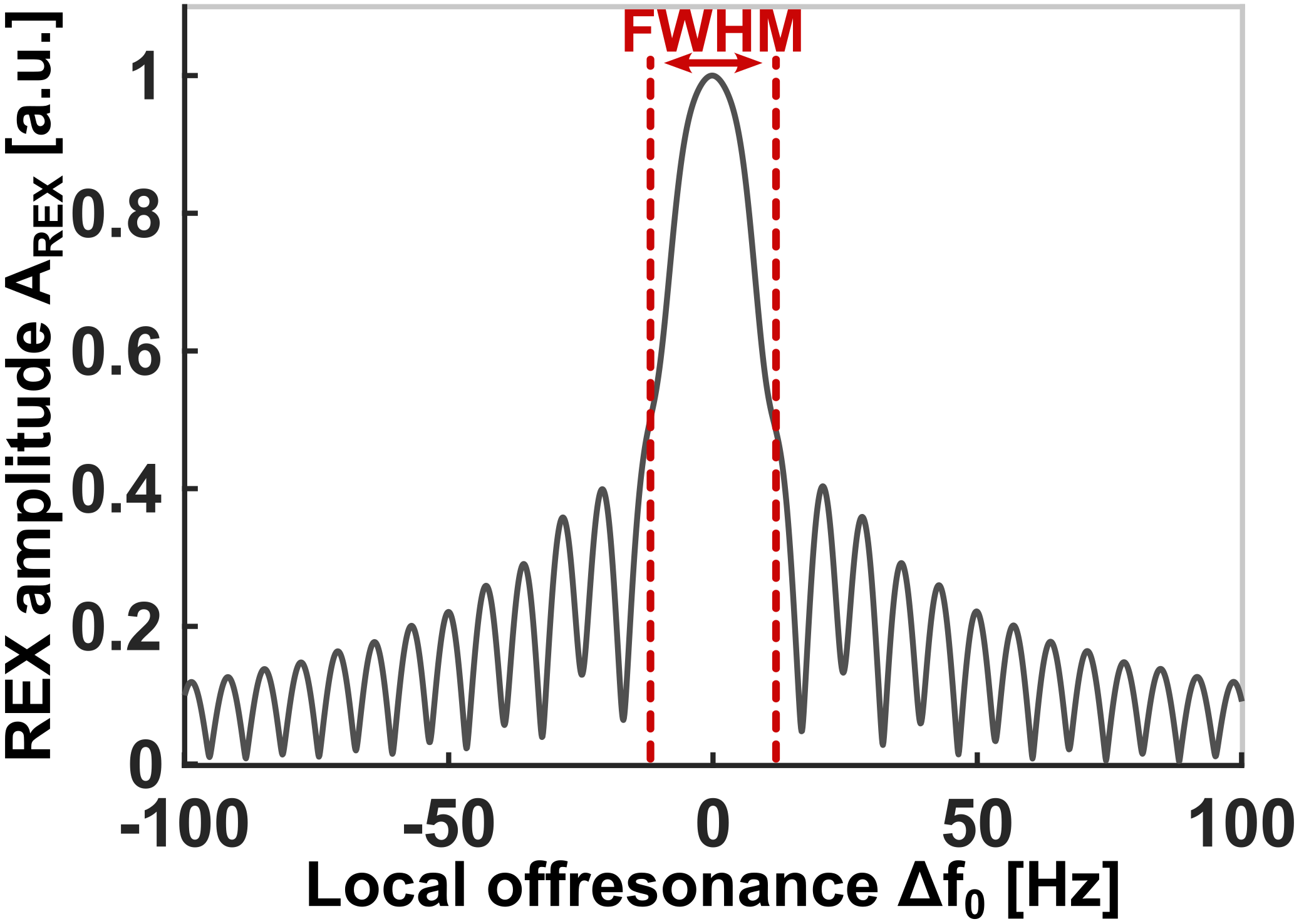}
\end{figure}

\newpage
\section*{References}
\textbf{[S1]} Duyn, J. H., Yang, Y., Frank, J. A. van der Veen, J. W. Simple correction method for k-space trajectory deviations in MRI. J. Magn. Reson. 1998;132(1), 150–153. doi: 10.1006/jmre.1998.1396
\\\textbf{[S2]} Schuenke P, Windschuh J, Roeloffs V, Ladd ME, Bachert P, Zaiss M. Simultaneous mapping of water shift and B1 (WASABI)-Application to field-Inhomogeneity correction of CEST MRI data. Magn Reson Med. 2017 Feb;77(2):571-580. doi: 10.1002/mrm.26133
\\\textbf{[S3]} Charagundla SR, Borthakur A, Leigh JS, Reddy R. Artifacts in T(1rho)-weighted imaging: correction with a self-compensating spin-locking pulse. J Magn Reson. 2003 May;162(1):113-21. doi: 10.1016/s1090-7807(02)00197-0
\\\textbf{[S4]} Witschey WR, Borthakur A, Elliott MA, et al. Compensation for spin-lock artifacts using an off-resonance rotary echo in T1rhooff-weighted imaging. Magn Reson Med. 2007;57(1):2-7.\\
doi:10.1002/mrm.21134
\\\textbf{[S5]} Gram M, Seethaler M, Gensler D, Oberberger J, Jakob PM, Nordbeck P. Balanced spin-lock preparation for B1 -insensitive and B0 -insensitive quantification of the rotating frame relaxation time T1$\rho$. Magn Reson Med. 2021 May;85(5):2771-2780. doi: 10.1002/mrm.28585
\\\textbf{[S6]} Hock M, Terekhov M, Stefanescu MR, et al. B0 shimming of the human heart at 7T. Magn Reson Med. 2021;85(1):182-196. doi:10.1002/mrm.28423
\\\textbf{[S7]} Schär M, Vonken EJ, Stuber M. Simultaneous B(0)- and B(1)+-map acquisition for fast localized shim, frequency, and RF power determination in the heart at 3 T. Magn Reson Med. 2010;63(2):419-426. doi: 10.1002/mrm.22234
\\\textbf{[S8]} Reeder SB, Faranesh AZ, Boxerman JL, McVeigh ER. In vivo measurement of T*2 and field inhomogeneity maps in the human heart at 1.5 T. Magn Reson Med. 1998;39(6):988-998.\\
doi:10.1002/mrm.1910390617
\\\textbf{[S9]} Noeske R, Seifert F, Rhein KH, Rinneberg H. Human cardiac imaging at 3 T using phased array coils. Magn Reson Med. 2000;44(6):978-982. doi: 10.1002/1522-2594(200012)44:6<978::aid-mrm22>3.0.co;2-9

\end{spacing}
\end{document}